# Practitioner Voices Summit:
# How Teachers Evaluate AI Tools through Deliberative Sensemaking

Dorottya Demszky, Christopher Mah, Helen Higgins

Stanford University

**Acknowledgements:** This study would not have been possible without the contributions of all sixty one participants (listed below). We are also grateful the hard work to members of the Stanford EduNLP lab and external organizations for supporting the planning, facilitation, and data collection from the summit, in alphabetical order: Jessica Ann, Liliana Deonizio, Amber Dhanani, Jeannette Garcia Coppersmith, Rebecca Hao, Isabelle Hau, Lucía Langlois, James Malamut, Rizwaan Malik, Judit Moschkovich, Alex Nam, Catherine O'Connor, Lena Phalen, Hannah Rosenstein, Mei Tan. We also thank Kaci Peel, Melissa Cromosini, Finn Laubscher and Poonam Saho for administrative and logistical support. We are also especially thankful our teacher advisory group of attendees who participated in pre-summit planning sessions: Chris Anderson, Michael Bostick, Shengjie Chen, Sarah Cook, Elizabeth Greer, Swati Jogdand, Zhengqing Li, Marlon Matilla, Julianna Messineo, Christine Persun, Al Rabanera, John Seelke, Rebecca Weaver, Catherine Wilson and to Sherenna Bates and Shelton Daal from Digital Promise for advising our co-design efforts. Thank you to Nikkie Zanevsky and Kysie Jensen for supporting this work at the Gates Foundation, and to David Gormley for assistance with data digitization. Summit participants, in addition to teacher advisory group members, included: Alejandra Abreu Avila, Patrice Allen, Michael Barrette, Rachel Beverly, Rachel Borden, Alisa Brown, Ashley Burch, Jessi Cummings, Jasmine Davis, Alexandria Donald, Whitney Doucette, Laura DuMont, Gwen Faulkner, Madison Gallagher, Jennifer Garst, Wilmer Granados, Olivia Hammer-Grant, Vanessa Hernandez, I Ling Hsiung, Geoffrey Kent, Moe Htet Kyaw, Celine Liu, Mark Lobaco, Dixie Lopez, Griselda Lopez-Mendez, Keely Machmer, Justice Martinez Gudorf, Elizabeth Mateo, Colleen McCarthy, Mitchelle McLeod, Abbie Meyer, Karen Monrreal Perez, Karie Mullassery, Valentina Nakic, Griselda Ramirez, Derek Sanders, Kavitha Satya-Mohandoss, Pamela Seda, Victoria Sepe, Carwai Seto, Tracey Shipman, Andres Soto, Gabriel Terrasa, Tamyra Walker, Olivia Wheeler, Marlon Williams, Xuan Xuan Zheng.

**ORCIDS:** D.D. (https://orcid.org/0000-0002-6759-9367), C.M. (https://orcid.org/0009-0004-6805-6876), H.H. (https://orcid.org/0009-0008-3222-9893)

**Ethics:** The study was approved by Stanford's Institutional Review Board (IRB #79890).

**Conflict of interest:** C.M., formerly at Stanford University, is now employed by Anthropic, an AI company. Anthropic had no role in the organization of the summit, and no Anthropic tools were used during the event. The authors declare no conflicts of interest.

**Data availability:** De-identified, original criteria are available at https://osf.io/bcpfv/overview?view_only=dd343da468f1481ab167d2544e23300c.

**Funding:** This work was supported by the Gates Foundation (grant #068816).

Correspondence concerning this article should be addressed to Dorottya (Dora) Demszky, 485 Lausen Mall, Stanford, CA 94305. Email: ddemszky@stanford.edu

**Author contributions.** Dorottya Demszky: Conceptualization, Methodology, Data curation, Investigation, Funding acquisition, Formal analysis, Writing – original draft, Writing – review & editing. Christopher Mah: Conceptualization, Methodology, Data curation, Investigation, Formal analysis, Writing – original draft, Writing – review & editing. Helen Higgins: Conceptualization, Methodology, Data curation, Investigation, Formal analysis, Project administration, Writing – original draft, Writing – review & editing.

## Abstract

Teachers face growing pressure to integrate AI tools into their classrooms, yet are rarely positioned as agentic decision-makers in this process. Understanding the criteria teachers use to evaluate AI tools, and the conditions that support such reasoning, is essential for responsible AI integration. We address this gap through a two-day national summit in which 61 U.S. K–12 mathematics educators developed personal rubrics for evaluating AI classroom tools. The summit was designed to support *deliberative sensemaking*, a process we conceptualize by integrating Technological Pedagogical Content Knowledge (TPACK) with deliberative agency. Teachers generated over 200 criteria — initial articulations spanning four higher-order themes (Practical, Equitable, Flexible, and Rigorous) — that addressed both AI outputs and the process of using AI. Criteria contained productive tensions (e.g., personalization versus fairness, adaptability versus efficiency), and the vast majority framed AI as an assistant rather than a coaching tool for professional learning. Analysis of surveys, interviews, and summit discussions revealed five mechanisms supporting deliberative sensemaking: time and space for deliberation, artifact-centered sensemaking, collaborative reflection through diverse viewpoints, knowledge-building, and psychological safety. Across these mechanisms, TPACK and agency operated in a mutually reinforcing cycle — knowledge-building enabled more grounded evaluative judgment, while the act of constructing criteria deepened teachers' understanding of tools. We discuss implications for edtech developers seeking practitioner input, school leaders making adoption decisions, educators and professional learning designers, and researchers working to elicit teachers' evaluative reasoning about rapidly evolving technologies.

**Practitioner Notes**

*What is already known about this topic*

- Research on AI use by teachers has largely drawn on Technology Acceptance Model (TAM) frameworks, foregrounding perceived usefulness and ease of use as drivers of adoption.
- Effective and responsible technology integration by teachers requires TPACK (integrated knowledge teachers need to teach effectively with technology), and is supported by agency (teachers' capacity to make principled professional choices).
- Hands-on engagement with AI tools and collaborative reflection with peers can support teachers' professional learning about AI.

*What this paper adds*

- Teachers' evaluative criteria for AI tools are multidimensional, spanning practicality, equity, flexibility, and rigor — and contain productive tensions such as personalization versus fairness and adaptability versus efficiency.
- The vast majority of criteria framed AI as an assistant executing teacher-defined goals; only 3% positioned AI as a coach for professional learning, suggesting a gap between current tool design and what research indicates is possible.
- A two-day summit designed around deliberative sensemaking supported teachers' evaluative reasoning through five mechanisms (time and space for deliberation, artifact-centered sensemaking, collaborative reflection through diverse viewpoints, knowledge-building, psychological safety), developing their TPACK and agency in a mutually reinforcing cycle.

*Implications for practice and/or policy*

- Edtech developers should engage diverse groups of teachers in hands-on, deliberative evaluation activities (not just surveys) to surface the criteria that matter for classroom use.
- Educators and professional learning designers can use collaborative, artifact-centered evaluation activities, such as rubric construction, as vehicles for building the very knowledge teachers need to make informed judgments about AI.
- School leaders should consult teachers' evaluative criteria when making AI adoption decisions alongside the compliance and scalability concerns that typically drive procurement.
- Short, intensive convenings offer a feasible complement to surveys and long-term co-design for both eliciting and supporting teachers' evaluative reasoning about rapidly evolving AI tools.

## Introduction

> "Design AI tools that honor human connection… and make space for bringing out creativity in human brains, instead of replacing it."
>
> - *Elementary math teacher, summit participant*

From U.S. federal policy to district-level decisions, the prevailing approach to AI in education positions teachers as recipients of tools and training rather than evaluators of them. A recent executive order, for instance, calls on the nation to "invest in our educators and equip them with the tools and knowledge to [...] utilize AI in their classrooms to improve educational outcomes" (Executive Order 14277, 2025). National organizations echo this language, promoting AI integration through toolkits and guidance frameworks (CoSN, 2026; TeachAI, 2025; US Department of Education, 2025). This top-down orientation extends to procurement: companies market AI products to districts, and administrators select and approve tools for teachers to use (CRPE, 2024). As Sarah, a participant in our study put it, "AI kind of breathes down all of our backs." Yet despite this push, as of 2025, only about a third of teachers reported using AI in their work on a weekly basis (Gallup, 2025). Rather than treating this gap as a problem of adoption, we frame it as reflecting teachers' ongoing judgment of whether and how new tools serve their students (Harvey et al., 2025). Given that teachers who want a primary role in technology decisions are largely shut out of them (Digital Promise, 2014; CRPE, 2024; Penuel et al., 2026), usage becomes one of the few sites where they can exercise professional judgment.

This judgment is an expression of teacher professional agency: the capacity to "make choices, take principled action, and enact change" in one's professional context (Anderson, 2010, p. 541; Molla & Nolan, 2020). Agency is not a fixed individual trait but is achieved in context, as teachers negotiate constraints including policy and leadership demands, institutional norms, resource limitations, and accountability pressures (Eteläpelto et al., 2014; Molla & Nolan, 2020). Supporting agency therefore means expanding teachers' opportunities to pursue what they value (Sen, 2012) by strengthening professional capabilities and improving the conditions under which they can exercise judgment (Molla & Nolan, 2020), so that they can in turn be partners in tool development (Penuel et al., 2026). Yet, the current landscape rarely provides such conditions. Teachers hold widely divergent views of AI shaped by their values, subject matter, student populations and prior experiences with technology. Any given school faculty might contain this full spectrum of viewpoints. Without structured opportunities to surface, examine, and negotiate these differences, adoption decisions default to top-down processes that bypass the reasoning this paper argues is essential.

From the perspective of agency, generative AI does not merely introduce new instructional possibilities; it also introduces new constraints and tensions that teachers must work through agentically. These constraints include opaque model behaviors, shifting district guidance and vendor policies, uncertainty about privacy and data governance, time costs of learning and monitoring tools, and the risk that tools subtly reconfigure instruction in ways that conflict with teachers' goals for mathematical sensemaking, equity, and student agency (Harvey et al., 2025).

A central way teachers can enact agency in this new landscape is by constructing and applying criteria for deciding which tools are appropriate, under what conditions and for which situations. These judgments go beyond a binary decision to adopt or reject an AI tool. They involve weighing multiple, sometimes competing considerations: Will this tool support mathematical reasoning or encourage cognitive offloading? Will it broaden access or widen inequities? Will it align with my instructional goals and constraints, or add to my workload?

This thought process requires what we call *deliberative sensemaking*: evaluative reasoning in which teachers draw on their values and professional knowledge to weigh trade-offs and arrive at context-sensitive judgments about AI tool use. It is *deliberative* because it demands the kind of sustained, effortful reflection central to deliberative agency, through which teachers surface values, question assumptions, and justify decisions (Kahneman, 2011; Molla & Nolan, 2020). It is *sensemaking* because it is grounded in the situated, integrated knowledge teachers bring to their practice — of content, pedagogy, students, and technology (Mishra & Koehler, 2006) — and facilitated by collaboration across perspectives. Supporting teachers' deliberative sensemaking, and understanding the criteria that emerge from it, can inform education technology design and adoption. This process is also valuable in its own right, since it is through deliberation that teachers develop the evaluative clarity needed to navigate AI in their own practice.

**Teacher adoption, attitudes and trust in AI**

A large body of research has examined whether teachers adopt digital tools and which factors shape their intentions and usage. Drawing on the influential Technology Acceptance Model (TAM), research has shown that perceived usefulness, perceived ease of use, and teacher attitudes are central drivers of teachers' intentions to use technology (Scherer et al., 2015, 2019), with external factors such as self-efficacy, subjective norms, and institutional support influencing use largely through their effects on these perceptions (Chien et al., 2014; Dogan et al., 2021).

More recent work extends this tradition to AI and large language models. Cross-national survey studies suggest that AI understanding and self-efficacy are strongly associated with perceiving more benefits, fewer concerns, and higher trust in AI in education, whereas demographic factors play a more limited role (Bergdahl & Sjöberg, 2025; Viberg et al., 2024). Other studies highlight the importance of credible information about AI and institutional support in shaping perceived usefulness and adoption intentions (Hazzan-Bishara et al., 2025), and document that teachers see substantial potential in AI tools but also worry about workload, job displacement, student dependence, and unintended classroom harms (Alwaqdani, 2025; Yin et al., 2025). However, this literature is largely structured around the question of whether teachers will adopt AI, rather than how they reason about what makes a tool appropriate for their practice.

Understanding this reasoning, which involves weighing pedagogical fit, student needs, ethical concerns, and institutional constraints simultaneously, requires looking at adoption through a lens of teacher agency and professional judgment.

**TPACK and teacher-centered design with AI**

Understanding how teachers exercise professional judgment, particularly with a technology as complex as generative AI, requires attention to the knowledge they draw on. The Technological Pedagogical and Content Knowledge (TPACK) framework defines good teaching with technology as an integrated, context-specific professional capability: “an emergent form of knowledge” that “goes beyond” content, pedagogy, and technology considered separately, requiring judgment rather than one-size-fits-all prescriptions (Mishra & Koehler, 2006). In the age of generative AI, developing TPACK becomes even more demanding, because teachers need specific forms of knowledge about tools that are “protean, opaque, and unstable” while also being “generative and social” (Mishra et al., 2023). For example, deciding whether to assign an LLM-based homework helper requires teachers to jointly weigh the capabilities and limitations of the tool itself (technological knowledge), approaches for supporting productive struggle rather than cognitive offloading (pedagogical and content knowledge), patterns of help-seeking and belonging in the classroom community (knowledge of students), and district guidance and access to tools at home (knowledge of institutional constraints). Exercising evaluative judgment about AI thus demands integrated professional knowledge that teachers must actively construct.

A growing body of work explores how professional learning experiences can help teachers build TPACK through case-based approaches combining instruction with hands-on practice (Ding et al., 2024), structured reflection scaffolded by peers and AI within communities of practice (Erbay-Çetinkaya, 2025), and identifying affordances and managing contradictions with AI tools (Tan et al., 2025). Across these studies, supporting teachers’ evaluative judgment appears to require not just information about AI, but conditions for sustained collaborative deliberation: time to reflect, diverse perspectives to challenge assumptions, and shared artifacts to anchor reasoning. However, fewer studies have characterized the practice-grounded criteria teachers construct when evaluating AI tools. With general-purpose LLMs now embedded across products and interfaces, it is especially important to understand the transferable criteria teachers use to evaluate AI-powered tools. This motivates our focus on eliciting teachers' criteria for evaluating AI tools broadly, and on examining how a structured, short convening can support the deliberative sensemaking needed to make those criteria explicit.

**The present study**

We address this gap by analyzing data from 61 K–12 educators who attended a two-day U.S. national in-person summit on AI in mathematics education. Our work makes two contributions. First, we characterize the criteria that educators prioritized when constructing rubrics for evaluating AI tools for their classrooms — criteria that extend beyond the dimensions typically foregrounded by technology adoption research. Second, we examine how a short, intensive convening—combining hands-on tool exploration, structured discussion, and diverse perspectives—can foster teachers' deliberative sensemaking about AI tools. In doing so, we position key activities from the summit as a methodological complement to surveys and long-term co-design: a format that can both elicit and support teachers' professional reasoning within a compressed time frame. Together, these contributions carry implications for how AI tools are designed, adopted, and introduced to teachers.

We address the following research questions:

1. **RQ1.** What criteria did K–12 mathematics educators articulate when constructing personal rubrics for evaluating AI tools for their classrooms?
2. **RQ2.** How did activities in a two-day in-person summit centered around the construction of criteria support educators'deliberative sensemaking?

## Methods

### Study Context and Design

We studied teachers' decision-making about AI in the context of a two-day summit titled "Practitioner Voices: Language Technologies in Math Education" held at [institution blinded for review] in June 2025. This fully funded event brought together 61 K–12 mathematics educators from across the U.S. — including classroom teachers, coaches, and school leaders — with intentionally diverse stances toward and experiences with AI. Our goal was to provide a platform for practitioners to inform research and development related to AI in mathematics education. Given overwhelming interest (300+ applications), participants were selected through an application process designed to maximize diversity in grade levels, roles, school contexts, and familiarity with and attitudes towards AI tools. Participants received travel support and continuing education units (CEUs). More details about the recruitment and selection process are included in Appendices A and B.

The summit was designed to support deliberative sensemaking. Figure 1 illustrates how design features mapped onto the two foundations of this construct, deliberative agency (Molla & Nolan, 2020) and TPACK (Mishra & Koehler, 2006). Our primary design commitment was to create conditions for deliberative agency rather than to develop TPACK directly, which requires more sustained, classroom-embedded engagement (Mishra et al., 2023). To support agency, we grounded activities in participants' values, fostered a curious, critical stance toward AI, and built in time for deliberation. To support TPACK, we recruited participants with diverse professional knowledge and experiences, briefly introduced how LLMs work and facilitated hands-on tool exploration grounded in classroom tasks. Collaboration across diverse perspectives connected (and reinforced) these two dimensions. This framework also guided our analysis of RQ2, where we examined how summit activities supported teachers' criteria development.

**Figure 1**

*Theoretical Lenses that Informed the Summit Design and Our Analyses*

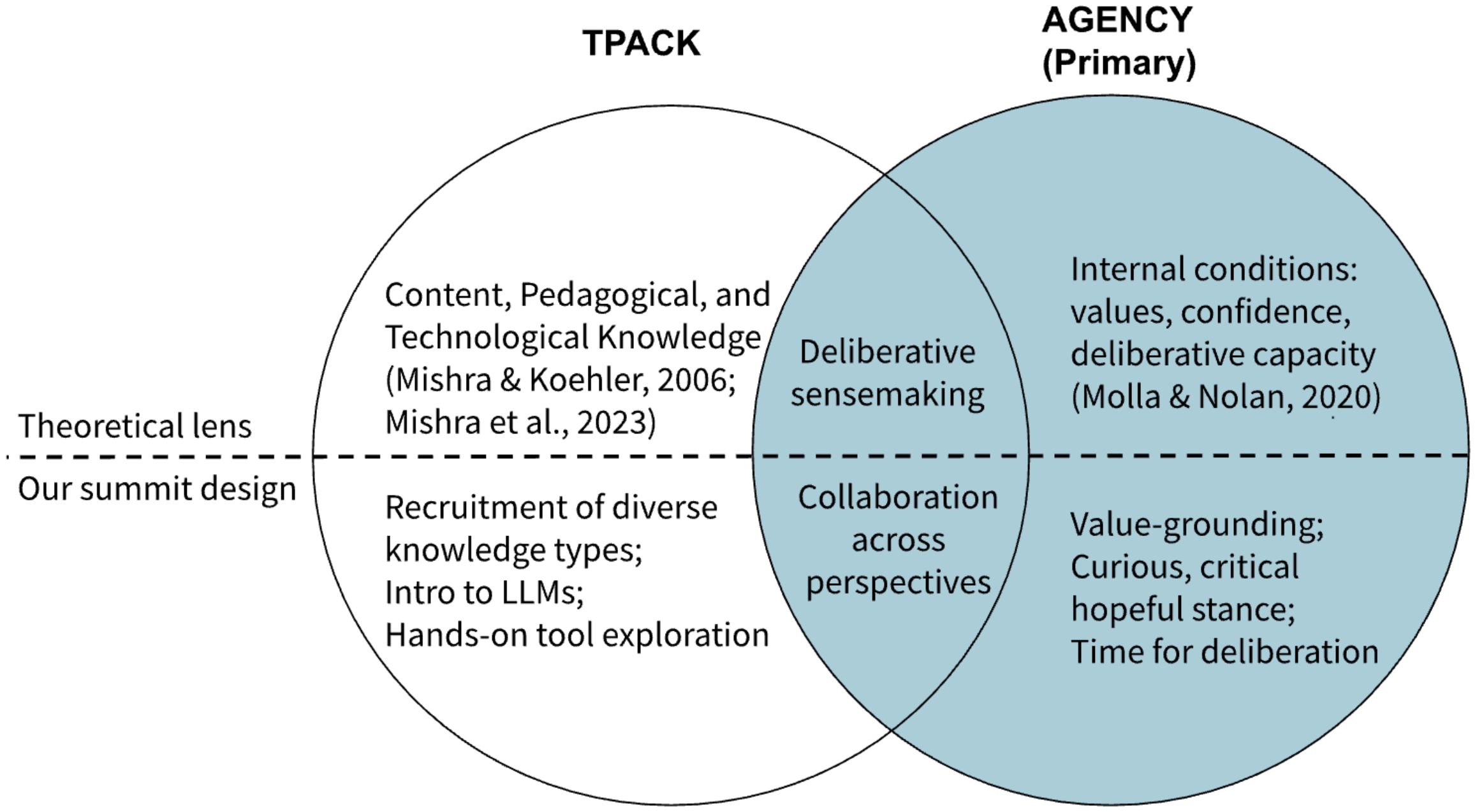

## Participants

Although research participation was voluntary and separate from summit attendance, all 61 summit participants opted in, consenting to audio recording of conversations and linkage of demographic data to their artifacts. Participants came from 22 states and represented a range of roles (50 classroom teachers, four school/district leaders, and seven instructional specialists or consultants), grade levels (elementary through high school), and years of classroom experience (Appendix Table B2). Some educators came from large districts in big cities; others from smaller districts in more rural areas. The majority of participants worked in schools serving predominantly low-income communities, with over half reporting school free/reduced lunch rates of 60% or higher. Fifty-six percent identified as people of color, and about a third were first-generation college graduates.

Most participants were already using AI tools in their classrooms, with about 10% reporting they used them "a lot," 35% "some," and 27% "a little," while 19% were planning to adopt them soon, indicating a slightly higher, but not too different from the national average of roughly 60–61% of teachers using AI in any capacity, as reported by Gallup (2025) and Education Week (2025). On a five-point skepticism scale, the mean score was 2.17, with most participants (74%) rating themselves a 1 or 2 (not skeptical), about 20% a 3, and only three rating themselves a 4, suggesting a generally receptive though not uncritical cohort.

## Pre-Summit Co-design

We hosted an array of co-design opportunities with participants in the months leading up to the summit. All confirmed attendees were invited to participate in synchronous design sprints, individual interviews or asynchronous short surveys to advise our planning and facilitation. Twenty participants responded to our co-design request and 13 participated in our interactive design sprints. While not a part of our data collection, the design sprints, interviews and surveys helped us refine our summit goals and develop a more effective prompt for our tool exploration session (Appendix C).

**Summit Procedures**

The summit included whole group sessions as well as breakout sessions that participants selected before arrival (see full agenda in Appendix D). Across two days, attendees spent five and a half hours in whole group sessions and three hours in their chosen breakout session. Breakout sessions included 14 to 16 people and explored specific AI applications (curriculum adaptation, feedback to students, feedback to teachers, analysis of student talk) in greater depth (see Appendix E for a description of breakouts, and any focal tools). In the present study, we only analyze data from whole group sessions, which were designed to build sequentially towards creating criteria. They began with *grounding*, moved into *exploration*, and culminated with *rubric development*.

*Grounding* (Day 1 morning). We invited a "curious, critical, and hopeful" stance and designed our first morning together to support all participants to feel included and valued (Appendix F). Our first small group conversations centered on values reflection through the card game Re-Ignite. Participants then connected their personal values to challenges they experience in their practice. Table groups then engaged in an AI or Not Card Sort, a task in which they categorized various forms of technology as AI or not and explained their reasoning, which offered an entry point for all experience levels in the discussion. Primed by these discussions, our final grounding activity to launch the summit included brief presentations about the history, definition, and modern context of Large Language Models (LLMs), after which, participants could ask researchers questions or write questions down to be answered later.

*Exploration* (Day 1 morning & afternoon). Next, we dove into tool exploration. In new small groups of similar roles and a variety of AI familiarity, participants shared their challenges and determined a shared scenario in which they might turn to an LLM (e.g. come up with a culturally responsive math word problem for students from Mexico). Groups then explored how ChatGPT might help them address their scenarios. They generated anchor charts outlining the evaluation criteria each group found meaningful at determining the quality of support offered by ChatGPT (Appendix G). From this shared criteria, educators reflected on personal criteria most important for them.

*Rubric Development* (Day 2 morning). We launched our second day by reflecting back patterns found in the criteria selected by participants on day one. We framed our rubric development session with the question: "How can we make criteria actionable so that they are helpful in determining if a tool is right for a situation?" Participants had ten minutes of drafting time, shared their draft with a partner, then revised their rubrics. We encouraged participants to consider their values, purpose, and audience, and to structure rubrics with specific criteria and descriptions of high and low performance (see Appendix H for instructions and Appendix I for example rubrics). We prioritized discussion time in table groups through several rounds of sharing their rubrics, encouraging participants to focus on the reflection process instead of the final product. This brevity was by design. Developing more detailed rubrics would require specifying the instructional context and task being evaluated, which the summit intentionally left open to capture breadth across diverse practitioners. The resulting rubrics are best understood as launching points that can seed more targeted engagement around specific scenarios.

**Data Collection and Processing**

We collected multi-modal data: audio/video recordings of selected sessions, surveys and interviews, and educator-created artifacts (e.g., handwritten notes and personal rubrics that articulated and ranked evaluation criteria for specific teaching scenarios).

*Survey Data.* Participants filled out a pre-summit survey that included questions about their background, school context, experience with and perspectives on AI, interests, goals for the summit (Appendix J). After the summit, participants completed a post-summit survey about their experience at the summit, with targeted questions on different activities (Appendix K).

*Recordings.* We recorded selected table groups during whole group sessions and breakout sessions to capture dialogue during activities.

*Rubrics.* Facilitators encouraged participants to surface criteria that mattered to them, name the rubric's intended purpose and audience, and frame language for readers less familiar with classroom contexts. We asked them to provide concise descriptors for the low and high ends of performance for each criterion (mid-points were not required). Rubrics were created either on paper or digitally by adding a slide to a shared deck that captured the participant's name, table number, purpose, audience, and a three-column table (criterion name; low; high). All submissions were collected and transferred into a shared spreadsheet. Our data includes rubrics for 55 teachers. Although these were intended to be personal rubrics, eight teachers completed them in groups (two groups of two, and one group of four). We disaggregated their data to the teacher-level for later analyzes. Participants were informed that their inputs would be combined and shared to inform AI tool builders and to support ongoing, practice-grounded conversations.

*Post-Summit Interviews.* Several months after the summit, we conducted thirty-minute, semi-structured follow-up interviews with seven of the 25 teachers who were sitting at recorded tables. We selected teachers who reported the maximum variation in feelings toward AI, as recorded on a Likert scale on a post-summit survey question. We started by asking what stuck with them. Then, using a stimulated recall approach (Calderhead, 1981), we showed them an excerpt of their group's discussion in which they were actively participating and asked them to reflect on 1) how others influenced their ideas in the discussion, and 2) how they intended to influence others' ideas. We also asked them to describe other professional learning around AI that they had experienced, as well as comment on how their experience would have been if they had completed the summit activities independently. We recorded and transcribed all interviews using Zoom or Google Meet and compensated the teachers we interviewed with a gift card. The interview protocol is included in Appendix L.

## Analysis

*RQ1: Criteria.*

Our analysis of the criteria focused on the 55 teacher-level rubrics, which together encompassed 211 individual criteria. Each rubric included between 2-6 criteria along with definitions of low and high values. Eight participants created their rubrics in groups, which we disaggregated to obtain individual-level data for analysis. We employed both inductive and deductive coding schemes. The raw criteria can be found on OSF[1].

First, three author-researchers independently open-coded a subset of rubrics, identifying recurring themes in criteria descriptions and their associated low/high definitions. When criteria were ambiguous (e.g., a single-word label like 'Adaptability'), we relied on context from the low/high values to interpret them. This process yielded a preliminary codebook of 18 criterion codes (Table 1).

Next, we grouped codes by whether they evaluated the AI *outputs* (the content students consume, such as formative assessments or chatbot conversations, 7 codes) or the *process* of using the AI tool (the teacher's experience, time, and workflow, 11 codes). This distinction helped separate teachers' considerations related to the pedagogical quality of the instructional materials from those related to the practical efficiency of the teacher's workflow.

Once we developed our preliminary codebook, we independently coded a calibration subset of twenty rubrics. A criterion could be assigned multiple codes. To maintain objectivity, the coding was conducted blind to the teachers' demographic and contextual characteristics. We reconciled disagreements and refined definitions, then assessed inter-rater reliability on an additional subset of ten rubrics, achieving strong agreement (average prevalence-bias adjusted kappa, PABAK = .86; range = .59–1.00), a measure chosen due to class imbalance across codes. The first author then coded the remaining dataset using the finalized codebook.

Building on prior work regarding AI in teacher professional learning (Jacobs et al., 2024; Li et al., 2025; Wang & Demszky, 2023), we also established an auxiliary variable to capture how the criteria positioned the role of the AI. We distinguished between an *assistant* role, which prioritizes the efficient execution of teacher-defined goals (e.g., drafting, retrieving, or formatting to speed up workflows), and a *coach* role, which aims to improve pedagogical practice through reasoning, justification, or constructive resistance to low-leverage requests. We coded criteria based on this variable in addition to the codes above by marking when a criterion explicitly positioned AI as a coach (PABAK = .93); this coding was quite conservative as we were aiming for high precision. Ambiguous examples were left unmarked.

Finally, to support interpretability, we aggregated codes into four higher-order themes: *Practical* (utility, feasibility, trustworthiness), *Rigorous* (learning quality, pedagogical soundness), *Equitable* (inclusion, access, fairness), and *Flexible* (teacher agency, versatile inputs/outputs). We assigned each criterion to the single theme that best captured its primary emphasis, while acknowledging that many criteria spanned multiple themes.

[1] https://osf.io/bcpfv/overview?view_only=dd343da468f1481ab167d2544e23300c

**Table 1**

*Codebook for Teachers' Criteria*

| Theme | Definition | Category |
|---|---|---|
| ***Output-focused (What do students see?)*** | | |
| Accessibility & UDL | WCAG/UDL, multilingual, reading levels, modality choice. | Equitable |
| Accuracy & Fidelity | Correctness, faithfulness to prompt, low hallucination. | Practical |
| Alignment to Learning Objectives | Matches the learning goals specified by the teacher. | Practical |
| Cultural & Community Relevance | Assets-based representation, local/contextual resonance. | Equitable |
| Encourages Productive Struggle | Supports productive struggle, talk moves, feedback quality, sequencing, "deeper" learning, supports academic rigor. | Rigorous |
| Engagement & Motivation | Captures interest without shortcutting learning; building student agency | Rigorous |
| Fairness & Bias | Minimizes stereotype reinforcement; parity across subgroups. | Equitable |
| ***Process-focused (What do teachers experience?)*** | | |
| Adaptability & Versatility | Flexible inputs/outputs, controls, modes, gives multiple options | Flexible |
| Contextual Awareness | Standards, curriculum, student data, classroom constraints. | Practical |
| Creativity & Novelty Support | Ideation beyond templates; divergent suggestions. | Flexible |
| Efficiency & Effort Reduction | Prep/time saved, fewer clicks, automation. | Practical |
| Pedagogy-Advancing Guidance | Points out higher-leverage alternatives, suggests moves to improve the teachers' instruction; role is always coach | Rigorous |
| Privacy & Data Governance | Collection, retention, FERPA/COPPA/IDEA alignment, on-device options. | Practical |
| Specificity & Actionability | Stepwise plans, feasible constraints, rubrics/checklists. | Practical |
| Teacher Agency & Control | Approvals, edit-ability, guardrails override, prompt steering, consistency. | Flexible |
| Teacher Voice & Style | Preserves tone, phrasing, and norms. | Flexible |
| Trustworthiness & Transparency | Cites sources, uncertainty/calibration, evaluation pedigree, provides rationale, reputable research-based responses. | Practical |
| Usability & Navigation | Information architecture, cognitive load, learnability. | Practical |

We also conducted an exploratory heterogeneity analysis examining whether teacher demographics (grade band, school poverty level, years of experience) predicted the distribution of criteria across themes; details and results are reported in Appendix M.

*RQ2: Activities that influenced criteria formulation.*

To analyze what activities influenced how teachers formulated their criteria, we drew on three sources: post-summit survey responses, observations from recorded summit conversations, and semi-structured interviews with select teachers. Participants were asked whether they preferred to be identified by their actual first name or a pseudonym; we use their preferred option throughout our reporting of RQ2.

To analyze post-summit survey responses, we followed qualitative coding methods from Miles and colleagues (2014). First, we selected a subset of six questions that related to activities that influenced how teachers' formed their criteria. We independently open-coded each teachers' responses with short, descriptive codes (e.g. "time to talk," "diverse perspectives," "hands-on"). Next, we discussed our codes, eliminated codes that we found irrelevant or least prominent, and grouped the remaining codes into three parent codes, each with 5-6 related child codes. The categories we created were:

1. *Structural*: This parent code related to structural features of the summit that were designed to facilitate hands-on collaboration. Child codes within this category included break-out sessions, hands-on activities, and overall organization.
2. *Content*: This parent code related to the content and activities that made up the summit curriculum. Child codes within this category included AI understanding, tool exploration, and rubric creation.
3. *Interpersonal*: This parent code related to relational components of the summit. Child codes within this category included collaboration, networking, and diverse perspectives.

We drafted a codebook with definitions and examples for each child code, shown in Appendix Table N1, then independently re-coded the data using the agreed-upon codes. Finally, we discussed and resolved discrepancies through a process of negotiated agreement (Campbell et al., 2013).

To analyze recorded conversations, we transcribed four groups' discussions (~4 hours), wrote analytic memos interpreting interactions through our codebook, and extracted relevant quotes. We then selected and interviewed seven of nine identified teachers using the methods described above. After each interview, we compiled our notes with takeaways and insightful quotes. For interview coding, we used a ChatGPT-in-the-loop workflow, motivated by recent evidence that LLMs can support reliable qualitative coding when paired with human verification (Goyanes et al., 2025; Theelen et al., 2024), as well as our own prior experience. We provided ChatGPT with our codebook, including parent codes, child codes, and definitions, and first prompted it to confirm its understanding of the scheme to reduce misunderstandings and unsupported inferences. We then uploaded each interview transcript and asked ChatGPT to indicate whether each child code was present and to extract supporting quotes for any code it identified. Finally, the interviewer for each transcript manually reviewed and validated the model's outputs, cross-checking against their interview notes to ensure the nature of the participants' sentiments were retained.

## Results

### RQ1: What criteria do teachers use for evaluating generative AI tools?

We start by reporting variation across teachers in terms of the criteria they highlighted, then move onto reporting variation across criteria. For between-criteria variations, we consider output vs. process focus, higher-order goals, and the implied role of AI (assistant vs. coach).

***Diversity in teachers' criteria.*** Teachers' criteria varied significantly. Figure 2 illustrates the frequency of criteria codes, calculated at the teacher level (representing 55 teachers, 297 teacher-criterion code combinations). To avoid skewing the distribution, each code was counted only once per teacher, even if a participant listed multiple criteria mapping to the same code; thus, a criterion code could occur at most 55 times. Teachers prioritized criteria related to trust, accuracy, and accessibility above all else. *Trustworthiness & Transparency* emerged as the most frequently cited dimension (27 instances), which focused on citing reliable sources and providing a rationale during the generation process (in contrast to *Accuracy*, that focused on output correctness). Teachers consistently expressed a desire for AI tools that not only provide citations but also admit their own limitations. For example, one teacher defined high performance as a tool that "openly says 'I don't know' somewhat frequently" and "asks good clarifying questions", whereas a low-performing tool would "present one ideas as the ONLY correct idea" (sic).

**Figure 2**

*The Distribution of Teacher-Level Codes for Criteria*

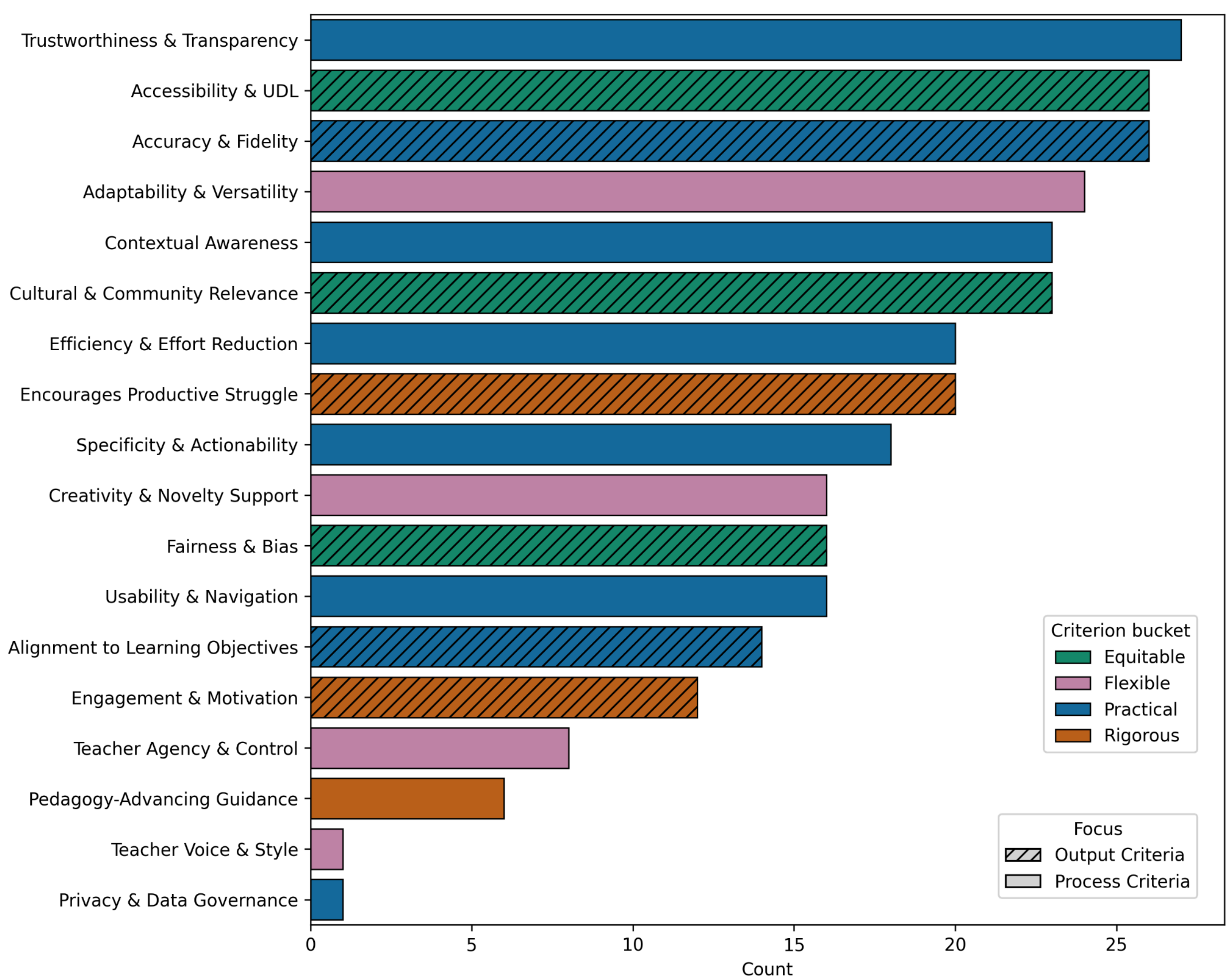

Note: Each criterion could be mapped to multiple codes, but we counted each code only once per teacher.

Following closely were *Accuracy & Fidelity* and *Accessibility & UDL* (26 instances each). Regarding accuracy, educators set high standards, with one rubric stipulating that a tool must generate "accurate math 100% of time" and be "aligned with the rigor and style of provided context 90% of the time." Regarding accessibility, teachers valued tools that could support diverse learners, describing ideal outputs as "low floor, high ceiling tasks that allow any level of students to engage" and tools that "prioritize access and success for all learners" and "actively reduce barriers and center marginalized voices."

Other highly prioritized criteria focused on the tool's ability to navigate the complexities of specific classrooms. *Adaptability & Versatility* (24 instances), *Contextual Awareness* (23 instances), and *Cultural & Community Relevance* (23 instances) were nearly as common as accuracy. Tools that score low on these criteria offer "limited options" or deliver a "word vomit to students." They also make "surface level attempts at connecting to students," such as the cliché of "teaching fractions... using a pizza." In contrast, educators sought tools that were technically robust, supporting "multiple modalities (text, voice, drawing)" while being able to "build upon past interactions" and "connect to learning management systems." Beyond technical integration and versatility, they prioritized cultural responsiveness, requesting "diverse examples" that are "sensitive to classroom settings" and capable of "honoring students' diverse cultures and languages... authentically." One teacher notably described high contextual awareness as a tool that "thinks like a teenager" to anticipate student sense-making.

Teachers considered efficiency gains, along with AI's potential to undermine learning. *Efficiency & Effort Reduction* and *Encourages Productive Struggle* appeared with equal frequency (20 instances each). Teachers valued tools where "the benefits outweigh the set up time" or that could produce quality outputs "more quickly than the teacher." At the same time, they penalized tools where students become "passive users" rather than doing "most of the cognitive work."

Beyond high-level goals, teachers focused on the practical usability of tools and their outputs, with criteria tied to *Specificity & Actionability* (18), *Usability & Navigation* (16), and *Alignment to Learning Objectives* (14). Their criteria downgraded "abstract ideas" in favor of "doable" results that offer "specific recommendations relevant to the current task", including ones that can be "turned into a PDF and printed with an answer key." They also penalized "clunky" interfaces, seeking "smooth, intuitive" experiences that fit seamlessly into their existing workflows and align with their "instructional intentions" and goals.

Teachers also valued AI that fosters equity and inspires innovation. *Fairness & Bias* and *Creativity & Novelty Support* were both cited 16 times, followed by *Engagement & Motivation* (12). Regarding equity, one rubric expressed concern about "blind spots," noting that a fair tool would have a "strong awareness of null curricula and the inherent bias of creators" while remaining "accurate." Teachers also valued AI as a partner for ideation, seeking "unique, 'out of the box'" suggestions and "fresh ideas in addition to tried and true standards," rather than repetitive or formulaic outputs.

Criteria explicitly discussing *Teacher Agency & Control* (8) were less frequent. While we identified specific requests for tools that "seek information" or honor "teacher judgment," agency was more often implied through other criteria expecting the tool to execute a variety of requests. *Teacher Voice & Style* appeared in only one criterion; this suggests that rather than seeking a digital substitute to mimic their persona, teachers desire a faithful assistant to support their work. Similarly, criteria related to *Pedagogy-Advancing Guidance* (6)—such as explicit requests for tools that push back against "bad" ideas and "enhance" instruction—were relatively rare (see discussion below on AI's role); though pedagogical quality was often implicit in criteria regarding engagement, alignment to classroom context and standards. Finally, *Privacy & Data Governance* appeared only once, likely because participants assumed systemic conditions like district approval and data security were already prerequisites.

Altogether, these criteria reflect diverse and high expectations that are characterized by inherent tensions. We explicate the trade-offs in greater detail in the Discussion.

***Evaluating the process vs the outputs of AI.*** In addition to analyzing variation across teachers, we analyzed variation in the focus of individual criteria, pooled across teachers (n=211). Criteria focused nearly equally on AI outputs (e.g., lesson plans, chat responses to students) and teachers' experience creating them (output: 127 criteria, process: 132 criteria). When evaluating outputs for students, teachers prioritized accuracy, accessibility and cultural relevance (Figure 1). When evaluating AI tools for their own use, they foregrounded criteria like trustworthiness, adaptability, contextual awareness, and efficiency. This indicates that for an AI tool to be adopted, it must not only generate high-quality learning materials but also respect the scarcity of a teacher's time. About a fourth of the criteria (n=53) touched on both the output and process, indicating that teachers perceived these dimensions as closely intertwined when assessing specific instructional goals.

***Goal of teachers' criteria.*** Figure 2 shows the distribution of criteria across the four higher-order categories. Practical criteria were most common (122 instances), nearly as frequent as the remaining three combined (Equitable: 65; Flexible: 46; Rigorous: 39). However, there was substantial overlap across categories: 70% of Flexible, 52% of Equitable, and 51% of Rigorous criteria were also labeled for at least one other category, indicating that the majority of criteria reflected multiple goals simultaneously.

***The role of AI in improving teachers' practice.*** The overwhelming majority of teachers positioned AI as an assistant that supported teacher-defined goals, which is consistent with their emphasis on *Practicality*. Only six criteria (3%) positioned the AI explicitly as a coach, expecting AI to improve teachers' instruction. One teacher indicated that they would "definitely use [AI if it] pushes back when a teacher prompts it with a bad/impossible idea." Another said that it should "provide valuable 'small change' practices that provide big results." One criteria named an example way in which AI might support teacher learning, by "providing some type of teacher notes to help teachers understand what about the selected [student] talk makes it a response worth sharing (how it connects to the big idea)".

Many criteria—even those not explicitly marked for the coach role—emphasized that AI must be grounded in 'sound pedagogy.' While we reserved the *coach* code for criteria that explicitly improved practice, the demand for pedagogical quality serves as a critical prerequisite for that role. Many criteria mentioned pedagogy supported by "research" or "best practices," anchored in sources ranging from academic research and established frameworks to practitioner experience. Still others located pedagogical expertise not within research or frameworks but within practitioners themselves. One teacher listed as their criteria, "Output aligned to [trained/experienced] educator expertise."

### RQ2: How did activities in a two-day in-person summit centered around the construction of criteria support educators' deliberative sensemaking?

Across the three data sources (post-summit survey responses, recorded table discussions, and follow-up interviews), we found convergent evidence that three key dimensions of summit activities fostered deliberative sensemaking: structural, design choices informing how the summit was organized and scaffolded; content, the curriculum presented; and interpersonal, elements that fostered connection and community. Because many interpersonal processes emerged as a result of structural and content features, we present this evidence through a set of higher-level mechanisms, which cut across data sources and codebook categories (Appendix Table N1). While each of the five mechanisms relates to each of the codes to a certain extent, Figure 3 illustrates the primary mapping between mechanisms and the three key codes (Structural, Content and Interpersonal).

**Figure 3**

*The Mapping of Deliberative Sensemaking Mechanisms onto Structural, Content and Interpersonal Dimensions of Summit Activities*

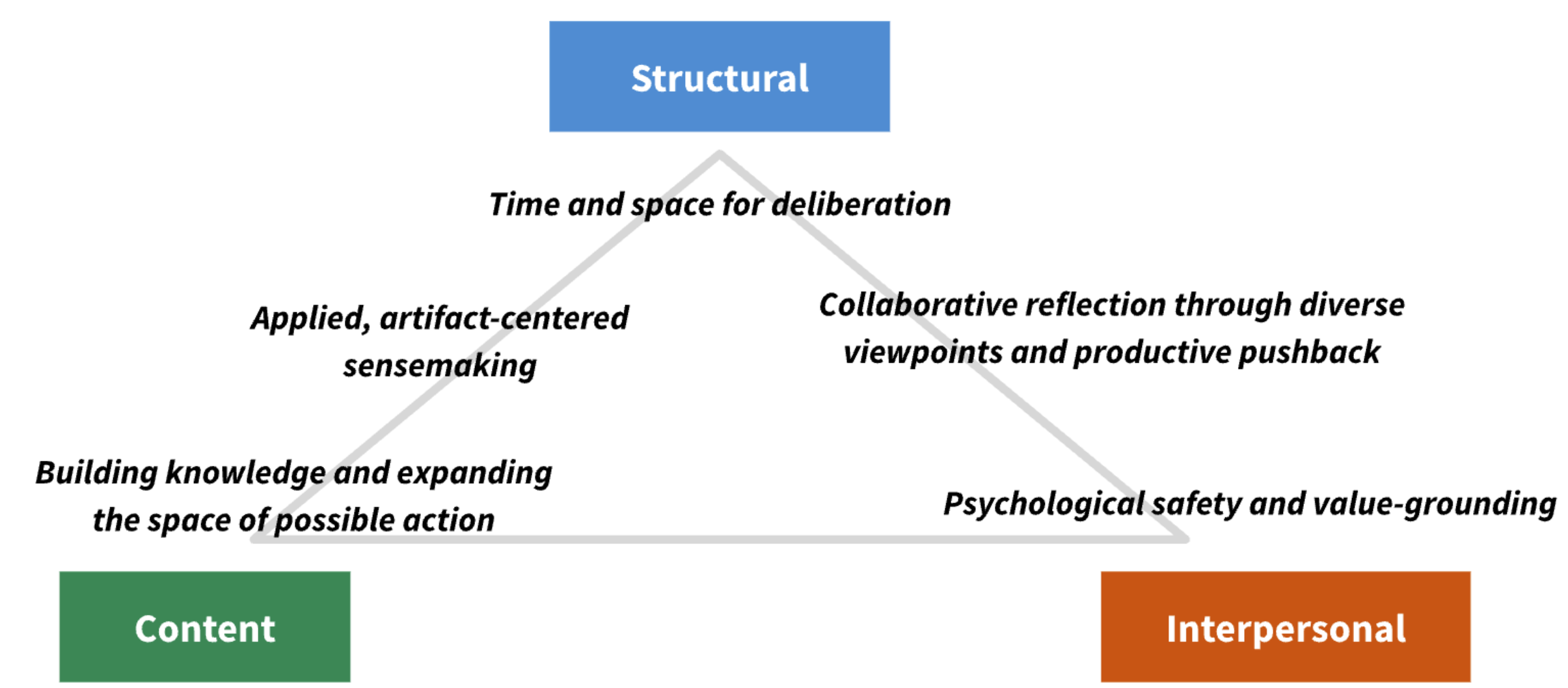

First we discuss findings for each mechanism, moving rowwise through the triangle in Figure 3. Survey counts refer to the number of survey responses coded with each code in our codebook. Counts for all codes are available in Appendix O. At the end, we provide a summary of overall feedback from the summit.

***Time and space for deliberation.*** The majority of teachers highlighted the provision of protected time and structured opportunities to talk, a foundational condition for deliberative agency (Molla & Nolan, 2020). High school math teacher Sarah described in her interview how the time for discussion deepened her engagement:

> *"I'm a math teacher. I think in bullet points, so I would have been done with that rubric in .6 seconds flat, and I would have been ready to do for the next thing [...] Having the whole conversations [...] with everybody having different usage, different entry points, different knowledge, it kind of opened it [...] too wide. [...] We actually had a harder time* *narrowing down the focus of what we were evaluating [...] because* ***the scope of AI and the scope of [...] how do we do this big, huge, nebulous thing, it was actually harder to refine. But I think that was the better challenge****."*

She acknowledges the immense challenge of navigating an open discussion with colleagues and recognizes the importance of that time and space when reckoning with something as complex as AI's integration into education.

In the survey, Time for Discussion was the most frequently cited structural feature (n = 28), with respondents emphasizing that the summit created "Time to talk and engage with peers and the researchers." Teachers saw the summit's overall organization (n = 9) as an enabling structure for sustained sensemaking. One survey respondent listed how the agenda created multiple venues for deliberation via breakout sessions, AI card sort, lunch time discussion and a "parking lot" of questions on the wall. Taken together, teachers' accounts suggest that the summit's temporal and organizational structure made sustained comparison, argumentation, and iterative refinement of evaluative criteria possible.

***Applied, artifact-centered sensemaking.*** Teachers consistently identified shared artifacts (tool outputs, prompting tasks, and draft rubrics) as anchors for deliberation. Hands-on activities in whole-group sessions and thematic breakout groups provided the structure for this anchoring. Eighth-grade math teacher Carwai reflected on how comparing outputs from a tool supported her group's conversation and her own learning:

*"Where my thinking changed a little bit was thinking about the way [others in my group] framed their prompts was very different than mine. Mine felt very clinical and very sort of straight to the point, [...] listening to [another participant's] prompts and the way she interacted with the AI gave her a different outcome. [...] I liked how diverse the experience was. [...] It reminded me that* ***how we choose to interact with this thing is also going to change what we get out of it****, which makes it interesting, because it's not just a machine that you press a button and the same thing comes out, but the choices you make also impact the output."*

Other teachers (n = 12) similarly noted that hands-on activities scaffolded their reflection. Andres from the feedback-focused breakout said, "Creating the feedback ourselves, then changing the feedback to what we would want, and sorting between multiple options of feedback was so thought provoking." The majority (n = 27) highlighted breakout sessions as spaces where abstract ideas became concrete; Alex recalled "being able to develop skills around crafting prompts that produce outputs more aligned to the things that you want." Several participants brought specific activities back to their schools: Carwai noted, "I brought [the AI or Not card sort] back to our school and we've actually done that. That has stuck with me," and Pamela described integrating CoTeach into a professional learning session she facilitated.

Across these activities, shared artifacts mediated most of the content teachers referenced: Tool Exploration (n = 23), Classroom Applications (n = 22), and Grounding via the Re-Ignite cards (n = 19). Most directly tied to deliberative agency, the majority of teachers referenced

Rubric/Criteria Development (n = 26), which the other activities built toward. Shengjie captured how the process deepened reflection on professional identity: "Developing criteria after exploring ChatGPT through scenarios that reflect my teaching beliefs, pedagogy, and student needs…deepened my reflection on the roles of AI and myself as a math teacher." Interviews suggest rubric development helped teachers shift from implicit preferences to explicit standards. Paule described how the language of "criteria for success" connected to her coaching practice and made the work click; Swati called the criteria work "eye-opening, [...] it really made me think about the ethics behind these tools," and notably reported that this shift transferred to her own classroom rubric-building; Alex emphasized that working through concrete examples surfaced specific risks — "what are some of the specific hallucinations that we're concerned about?" Together, these accounts suggest that artifact-centered activities supported deliberative sensemaking by helping educators build integrated TPACK while simultaneously articulating expectations, anticipating failure modes, and translating those judgments into more intentional tool use.

***Collaborative reflection through diverse viewpoints and productive pushback.*** A central mechanism through which the summit supported educators' deliberative sensemaking was collaborative reflection—spaces where teachers could raise, debate, and refine ideas with colleagues (Molla & Nolan, 2020). In the interview, Paule, an instructional coach, described a moment where a strongly anti-AI participant gradually revised his stance through group discussion, attributing this shift to the vulnerability and curiosity the group had established:

> *"We all just started going around as we're sharing our values from before…we started thinking about ways where [...] you've explained one side of the coin, what could the other side of the coin be with what AI is capable of […] Through those conversations of us blowing each other's minds […] by the end of that conversation, he's like, all right, well, I might not be a negative five [...] maybe a two or three [...] I think it was really cool how vulnerable we were in that card activity [...] with that foundation of, wow, these are real people [...] people that want to see students succeed [...] we all allowed ourselves to just show up vulnerably and with curiosity. And I think as he started to let his guard down, I think he became just more curious, more maybe open-minded."*

She contrasted this with working alone, which would have been "through one lens," calling the summit's diversity of perspectives irreplaceable: "it would have been a robbed experience if I did it by myself."

Nearly all (46 out of 48) survey responses explicitly referenced collaboration or connection as the "most valuable component" of the summit. The interpersonal codes most directly related to this mechanism were Diverse Perspectives (n = 25), Collaboration (n = 12), and Networking (n = 16). Teachers described how disagreement and heterogeneity made deliberation more rigorous rather than polarizing: Miguel emphasized "a lot of people with diverse perspective with whom I can disagree" as a route to "build[ing] consensus and a better understanding of our own practices," avoiding "a polarizing bubble."

Carwai's quote above described a different kind of shift, not in stance, but in understanding of the tool itself. Comparing his own "very clinical" prompting style to a tablemate's more creative, coaching-oriented approach supported his intentionality. Other interviewees echoed the idea that exposure to different kinds of experience, roles, and skepticism sharpened their own thinking. Alex noted that encountering skepticism "in different ways [...] definitely increases the level of the expectations." Pamela described how working alongside teachers with limited AI experience reoriented her attention toward communication and support: "being exposed to people in that group who had limited experience with AI really motivated me to think about how do I communicate, and how do I support teachers who have limited experience with AI." Miguel similarly emphasized how interacting with administrators and instructional coaches pushed his thinking from individual classroom use to system-wide possibilities for AI-supported collaboration.

Two recorded table conversations during whole group time illustrate these mechanisms in action. The first one (Table 2) centers on a skeptical participant (John) who argues that ChatGPT-generated word problems are not useful for generating culturally responsive word problems because they lack grounding and take too much effort. In response, peers push him to clarify what "grounded" should mean and how much burden should fall on the user. For instance, when John objects that a scenario is "made up," Pamela reframes the critique—"But most math problems are made up" (line 2)—prompting laughter and forcing John to specify that his concern is not fabricated numbers per se, but fabricated numbers when real data exists and should be retrievable. Then, Sarah offers a concrete tactic (line 4), John tries it, and he refines his criteria again: he wants not just better answers, but answers that provide credible sources by default (line 5). Sarah adds a second layer of pushback—explicitly aligning as an AI skeptic (line 6) while arguing that even humans require follow-up questions, again pressing John to separate ideological opposition from actionable evaluation criteria. This episode shows deliberative agency as a socially mediated refinement process: peers challenge assumptions, propose tests, and help surface more precise criteria (e.g., default sourcing, transparency, and user effort). See full excerpt in Appendix P.

***Table 2***

Excerpt of Group Conversation from Table I

| line | speaker | text |
|---|---|---|
| 1 | John | I looked at a prompt related to social justice math in Mexico for systems of equations, and [ChatGPT] gave me a basic scenario…I said, can you make it a little bit more complicated? And it made it a little more complicated. Then I even tried to do something with reparations and it gave two options or scenarios but **I want actual data. There's no sources here. Where's the data coming from? …**I'll just give you an example, right? In the problem scenario about Mexico, using systems of equations, it talks about fair wages. And then it says, you know, Anna worked 40 hours and earned blank pesos. Jorge worked 35 hours and earned blank pesos. You |
| | | overhear that Jorge may be earning more for hours because he's male, you want to figure out the hourly wage like- |
| | Pamela | It's all made up. |
| | John | Right, that's what I'm saying, so like- |
| 2 | Pamela | **But most math problems are made up** [laughter] |
| 3 | John | No, no, but there is actual data that you can access this information from about males and females working in these countries. And so I think that- |
| 4 | Sarah | Could you ask it like, what is the actual data where did you get your statistics from? **I know sometimes AI lies to you straight up, but-** |
| *Discussion continues as John tries re-prompting ChatGPT in this manner.* | | |
| 5 | John | **So, I admit when you led me, Sarah, to ask that question, it did give me a little bit more information, right?** It has data informed inspiration, but not direct resources. It gives me OECD, gender wage gap data, Mexican government labor ministry, things like that. **I would hope that I wouldn't have to ask multiple times for that. Like that should come off right away, to be able to help.** |
| 6 | Sarah | Not that I'm defending AI though, but that is how it would work with a human though, right? |

Another excerpt (Appendix P) illustrates a different interactional pattern: rather than pushing back, teachers build on each other's strategies. By comparing a highly specified lesson-creation prompt with an open-ended goal-focused prompt, the group converges on the principle that prompt specificity shapes whether AI functions as a lesson-writing assistant or a broader brainstorming partner — then extends this reasoning to instructional coaching. Here, collaborative reflection expands the space of possible action while simultaneously sharpening criteria for quality (fit to purpose, pedagogical leverage, and the teacher's ongoing role in steering outputs).

***Building knowledge and expanding the space of possible action.*** Molla and Nolan (2020) observe that deliberation "opens up an opportunity for learning how and why [teachers'] taken-for-granted assumptions and perceptions might constrain their ability to facilitate young children's learning." (p. 74) Laura, an elementary math teacher, captured this trajectory in the survey:

> *"This summit made me feel confident that I could try something new that I was intimidated by (ChatGPT!). I appreciated the way that AI was explained on the first day because it made me feel like I understood more about the various types of tools available and how they came to be. Before that understanding, I was honestly terrified to take a* *chance on an unknown tool.  Now I feel like I can "play around" with tools in order to see what fits my classroom and the needs of my students."*

Laura's shift from fear to informed confidence illustrates how knowledge-building can expand the space of possible action. Teachers referenced summit components that built their TPACK: Understanding of AI (n = 23), Tool Exploration (n = 23), and Gaining Confidence/Intentionality (n = 10). Several described epistemic shifts in how they related to AI tools. Carwai, for instance, arrived assuming "you didn't need to know facts and have that kind of…expertise necessarily, because the AI could support that," but came away convinced: "you actually need to be more of an expert to use this tool. Because you need to know where it goes wrong…if you really don't understand it, you are then perpetuating all sorts of errors." A survey respondent articulated a similar realization: "the AI is only as strong/useful as the educator themselves. At the core of this practice, there should be an educator or practitioner who is very aware and intentional about their practice and the use of the tool."

For many, these shifts translated into motivation to explore new possibilities with AI. In interview, Paule reported leveraging "some type of AI almost every single day" and co-facilitating an AI-focused PD for her staff; Swati described returning home with "inspiration and a motivation…to explore AI more"; and Miguel shifted from viewing AI as an individual supplement to envisioning cross-disciplinary collaborative structures among teachers. Across these accounts, knowledge-building and confidence operated as mutually reinforcing: deeper understanding of AI's capabilities and limitations gave teachers more precise criteria, which in turn gave them the confidence to act on those criteria in their own contexts.

***Psychological safety and value-grounding.*** Deliberative agency involves questioning practice "against the backdrop of [teachers'] own values and beliefs" and reassessing assumptions through critical reflection and openness to collegial critique (Molla & Nolan, 2020, p. 74). Fifth grade math teacher I Ling captured in her survey response how the value-grounding activities at the summit shaped her approach to AI:

*"One key takeaway from the summit is the importance of grounding my approach to using AI in the classroom in my personal values, such as kindness and integrity. As I consider integrating AI into my math teaching, I realize that these values should guide how I select and use technology, ensuring that AI tools are used to support all students equitably, foster a supportive classroom environment, and maintain trust between teacher and students. For example, kindness means using AI to provide encouragement and personalized support, while integrity involves being transparent about how AI is used and ensuring student data is handled responsibly. Ultimately,* ***centering my values helps me approach the challenges and opportunities of AI in a way that prioritizes student well-being, ethical practice, and inclusive learning for every student in my classroom****."*

Survey codes most closely aligned with this mechanism were Grounding (n = 19), Feeling Valued/Respected (n = 8), and Gained Confidence/Intentionality (n = 10). Interviewees described how grounding in values supported their evaluative clarity. Carwai noted that "understanding my values and what I care about has become [...] much easier for me to [...] decide, 'Oh yeah, that's useful. That's not.'" Swati similarly emphasized intentionality — treating AI "as a catalyst and not… a driver," insisting, "I don't want to stop thinking."

Values-based activities also enabled vulnerability, which participants described as a prerequisite for learning and perspective change. Paule, in the quote above, describes how this climate helped a skeptical participant "let his guard down" and become open-minded. Alex described a complementary outcome: encountering a range of stances "prepared me for other conversations" where they could validate others' positions while still inviting more careful consideration. Shengjie echoed feeling respected as contributors to sensemaking: "Our ideas and voices were truly well-listened to and valued there." Swati appreciated that the summit centered teacher experience, taking time to "hear what teachers think before" tool development. In combination, these accounts suggest that the summit's supportive atmosphere — grounded in values, collegiality, and respect — helped teachers share uncertainty, articulate boundaries, and engage in the kind of reflective, criteria-based reasoning that characterizes deliberative agency.

***Overall feedback.*** Post-summit survey responses (n = 54, 89% response rate) were overwhelmingly positive: all respondents rated the summit as productive or very productive (M = 4.57/5), all expressed interest in attending future events, and 59% volunteered as community ambassadors. Teachers' perceptions of AI's impact were cautiously optimistic, with 61% rating it positive and none negative (full results in Appendix Q). The most common requests for improvement concerned time pressure around the rubric activity and the openness of the criteria-development task, which some participants found difficult to navigate without more scaffolding.

## Discussion

This study set out to understand the practice-grounded criteria K–12 mathematics educators construct when evaluating AI tools for their classrooms, and to examine how a structured, short convening can support the deliberative sensemaking needed to make those criteria explicit. Our findings offer two main contributions. First, the criteria teachers articulated reveal considerations that add depth to our understanding of what shapes their decisions to adopt or reject a tool. Second, we surface mechanisms through which a short, structured convening can support teachers in deliberative sensemaking.

### What teachers look for in AI tools

***Criteria reveal teachers' multidimensional evaluation.*** The criteria teachers constructed add depth to the dimensions foregrounded by technology acceptance research, namely perceived usefulness and perceived ease of use (Scherer et al., 2019). Rather than any single concern dominating, teachers simultaneously weighed trustworthiness, pedagogical rigor, cultural relevance, and accessibility — and evaluated both AI outputs and the process of producing them. A tool that generates strong lesson plans but requires extensive re-prompting or offers no rationale may still fail their evaluation. These diverse, multidimensional criteria indicate that teachers are already making integrated, context-sensitive judgments about AI — the kind of reasoning that TPACK characterizes as an emergent form of professional knowledge (Mishra & Koehler, 2006). Understanding how teachers resolve these multiple, competing considerations can extend technology adoption frameworks with a focus on general perceptions towards situated evaluative reasoning.

***Educative tools?*** Only six (3%) of criteria positioned AI as a coach that could improve teachers' practice; the vast majority framed AI as an assistant executing teacher-defined goals. This pattern is consistent with current product designs that emphasize productivity. But it also suggests a gap between how teachers currently frame AI's role and what research on educative tools (Davis & Krajcik, 2005) indicates is possible — namely, that AI can serve not just as an assistant but as a catalyst for professional learning, supporting reflection, feedback, and pedagogical growth (Demszky et al., 2025; Erbay-Çetinkaya, 2025; Li et al., 2025; Malamut et al., 2025). Some participants began to articulate this possibility, describing AI as a "highly reflective teacher tool" or a "scaffold [...] until you develop that muscle to do it yourself." Yet others raised a counterpoint that outsourcing the generative work of planning and adapting instruction may erode the very expertise needed to evaluate AI's output ("if you don't do it yourself ever, you don't know how to be a judge of quality"). This worry about expertise erosion echoes long-standing debates in HCI about automation and skill degradation (Cukurova, 2026; Parasuraman, 2000), and may apply more acutely to tasks like lesson planning than to reflective tasks like analyzing one's own discourse. The field needs to pay more attention to what automation enables–or limits–in teachers' professional learning.

***Productive tensions in teachers' criteria.*** Teachers' criteria contained productive contradictions that mirror genuine dilemmas in teaching with AI. Three tensions stood out. First, *personalization versus fairness*: teachers wanted tools that tailor content to individual students' needs and cultural contexts, but also worried about bias and equitable treatment across subgroups — goals that can be at odds when greater customization makes "fairness" harder to define and verify. Second, *adaptability versus efficiency*: some teachers prioritized extensive control over AI outputs — seeking tools that support multiple modalities and respond to iterative refinement — while others foregrounded speed, valuing tools that produce quality outputs "more quickly than the teacher." This reflects a genuine trade-off, since a tool that lets teachers iteratively refine outputs across modalities demands more time and effort than one that delivers a ready-to-use lesson plan in a single prompt. Third, *aspirational standards versus current capabilities*: teachers described tools that would deeply understand curriculum scope, sequence, and depth of knowledge while also being sensitive to their specific classroom context. Yet this vision runs up against not just the limitations of current AI, but the boundaries of what the learning sciences themselves can specify — how to simultaneously optimize for personalization and fairness, or creativity and curricular alignment, remain open questions in education. Because the way these tensions are resolved depends on context that only the teacher fully knows, supporting teachers' capacity to reason through these trade-offs is essential for responsible AI integration.

**Mechanisms for deliberative sensemaking**

***Supporting deliberative sensemaking about AI.*** The summit's overwhelmingly positive reception suggests that educators have a strong, largely unmet need for structured opportunities to reason collectively about AI. The most common request for improvement was more time: for rubric development, for discussion, and for hands-on exploration. The responses signal that deliberative sensemaking about AI requires protected time, diverse interlocutors, and concrete artifacts to reason with. These conditions are rarely available in teachers' day-to-day professional lives, where AI-related decisions are often made under time pressure or delegated to administrators (CRPE, 2024).

***Mechanisms of deliberative sensemaking and the relationship between TPACK and agency.*** The five mechanisms that emerged from our analysis (time and space for deliberation, artifact-centered sensemaking, collaborative reflection through diverse viewpoints, knowledge-building, and psychological safety) provide empirical grounding for the framework in Figure 1. These mechanisms connected the two foundations of deliberative sensemaking, TPACK and agency, in mutually reinforcing ways. Knowledge-building activities (learning how LLMs work, exploring tools hands-on) gave teachers the technological and pedagogical content knowledge needed to construct meaningful criteria, while the act of constructing criteria — an exercise of deliberative agency — in turn deepened teachers' understanding of tools, surfacing limitations and affordances they had not previously considered. This bidirectional relationship, with TPACK enabling more grounded agency, and with agency generating deeper TPACK, extends prior work that has largely treated these as parallel constructs (Mishra & Koehler, 2006; Molla & Nolan, 2020). Our findings suggest that knowledge-building and evaluative judgment should be interleaved rather than sequenced, so that each strengthens the other through iterative cycles of exploration, reflection, and criteria refinement.

***Designing for openness and scaffolding.*** A productive tension that ran through the summit's design was between the deliberate openness that supported sensemaking and participants' desire for more structure. This tension is inherent to designing for deliberative sensemaking. Open-ended tasks are what allow teachers to surface their own values and construct criteria grounded in their specific contexts while being influenced by diverse perspectives around them— prescribing the outcome would have undermined the very agency the summit sought to support. At the same time, participants' readiness for that expansive thinking likely varied with the extent to which they had previously reflected on AI's role in their practice. The challenge for future designs is to provide enough scaffolding to make the task accessible without narrowing it so much that it forecloses the diversity of perspectives. Our data suggest that iterative rubric development as knowledge develops may help balance these demands, providing more scaffolding while preserving the open-endedness that supports genuine deliberation.

## Implications

***For edtech developers.*** Our findings corroborate prior work on practitioner-centered design (e.g., Garreta-Domingo et al., 2018; Penuel et al., 2022, 2026) that gathering meaningful teacher input on AI tools requires more than post-hoc feedback surveys or small convenience-sampled focus groups. The wide variation in teachers' criteria means that any narrow sample risks hearing only a slice of the evaluative landscape. Developers seeking teacher input should recruit purposefully for diversity of role, school context, experience level, and stance toward AI, and should use methods that elicit reasoning, not just ratings. Our richest data (surfacing tensions and trade-offs) emerged when teachers were in conversation with one another, grounded in actually using the tools. While multi-year co-design partnerships may not be feasible for smaller edtech organizations with limited resources, our findings suggest that a short, structured convening (analogous to a design sprint (Knapp et al., 2016)) can be an efficient way to surface diverse teacher perspectives, build trust, and energize educators as partners in the development process. Such a convening can then serve as a launchpad for more sustained engagement through smaller focus groups or interviews over time.

***For educators and professional learning designers.*** Similarly to prior work, we find that hands-on experience shapes teachers' attitudes towards AI (Viberg et al., 2024), particularly when embedded in collaborative learning communities (Tan et al., 2025). Participants who arrived skeptical or unfamiliar reported epistemic shifts, from assuming AI reduced the need for expertise to concluding the opposite, or from fearing the technology to feeling confident enough to "play around" with tools to find what fits their classroom. These shifts were produced by the combination of using tools, articulating criteria, and debating trade-offs with peers. They suggest that educators stand to gain from seeking out or creating such opportunities, whether in formal professional learning settings or informal communities of practice. Although the summit was not designed as professional development, teacher learning emerged as a meaningful byproduct. Our findings corroborate prior showing that combining direct instruction (i.e. our brief intro to LLMs) with collaborative problem-solving around specific cases supports teachers' AI literacy (Ding et al., 2024). They add that evaluative activities themselves can serve as a vehicle for building the very knowledge they draw on. Because teachers within any school are likely to hold divergent views of AI, even a short convening that brings them together with psychological safety and concrete artifacts to reason with can catalyze this process. Such convenings also present an opportunity to help teachers envision AI not only as an assistant but as a tool that can support their own pedagogical growth.

***For school leaders.*** The criteria teachers articulated in this study offer a window into what practitioners actually look for in AI tools, which may not always match what administrators may prioritize in procurement decisions (Yin et al., 2025). Privacy and data governance, for instance, rarely appeared — likely assumed to be district-level prerequisites — while teachers focused on pedagogical fit, cultural responsiveness, and trustworthiness. This division of evaluative labor means that if districts adopt tools based primarily on compliance, cost, and scalability without consulting teachers criteria, they risk selecting tools that meet institutional requirements but fail classroom use (lacking in practicality, pedagogical depth and equity orientation). Our findings suggest that school leaders should study and incorporate teachers' criteria before committing to AI tools, and should invest in opportunities that support deliberative sensemaking in communities of practice rather than one-off training sessions. The core conditions for productive deliberation we identify (e.g. protected time, diverse perspectives, hands-on exploration) are portable to far less resource-intensive settings, such as a school-level professional learning day.

***For researchers.*** We framed the summit as a methodological complement to surveys and long-term co-design: a format that can both elicit and support teachers' evaluative reasoning within a compressed time frame. Our findings bear this out. The summit produced richer, more nuanced data than closed-ended surveys typically yield, because teachers generated criteria in their own terms, debated their meaning, and situated them in specific instructional scenarios. At the same time, the two-day format was more nimble than longitudinal co-design partnerships, making it feasible to convene a nationally diverse sample without the sustained institutional commitments. This combination makes intensive convenings a particularly promising format for research on AI in education, where the rapid pace of technological change can exceed the timelines of traditional longitudinal studies (though sustained partnership remains essential for understanding how criteria evolve over time with use).

***Limitations and future directions.*** Our findings should be interpreted in light of several limitations. The summit was a two-day convening, and while this format enabled productive reasoning, it captured teachers' evaluation at a single point in time. Criteria developed over two days are necessarily initial articulations; with more time and sustained tool use, teachers might refine, expand, or fundamentally revise their rubrics. Participants' own feedback corroborated this, several noting the rubric development felt rushed. Accordingly, researchers and practitioners drawing on these criteria should treat them as a starting framework, and refine them through engagement with the specific contexts, tasks, and practitioners where tools will be used.

Although we sought to be intentional about recruiting AI skeptics, too, our participant pool likely overrepresents teachers already engaged with or curious about AI, and the findings may not generalize to other subject areas, grade bands, or national contexts. Additionally, while the criteria spanned a wide range of concerns, the near-absence of the "coach" framing (3%) suggests that the summit design may not have sufficiently scaffolded teachers to envision AI roles beyond current product paradigms.

A longitudinal follow-up could track how teachers' criteria evolve with sustained AI use and whether the tensions we identified (e.g., personalization vs. fairness, adaptability vs. efficiency) resolve, deepen, or shift over time. Expanding to other disciplines and international contexts would test the transferability of the criteria themes and higher-order categories we identified.  Future designs could also further scaffold the rubric-building activity, prompting teachers to articulate criteria for AI as a tool for their own professional learning alongside criteria for AI as an assistant.

Perhaps most importantly, future convenings could bring edtech developers into the process — not as presenters but as listeners — to test whether direct engagement with practitioners' evaluative reasoning changes how tools are designed. Such three-way partnerships among practitioners, researchers, and developers (cf. Luckin & Cukurova, 2019) could move the field from documenting what teachers value toward building tools and professional learning structures that are responsive to it, though they would require ensuring that practitioner voices remain central rather than being channeled toward corporate priorities.

# Appendix A
# Summit Recruitment Materials

## Recruitment Flier and Website

**Practitioner Voices: Language Technologies in Math Education**
**An Educator Summit at [institution blinded for review] | June 23-24**

The [lab name blinded for review] invites you to apply for an innovative summit where educators will help shape the future of language technologies in mathematics classrooms. Insights from participants will directly inform research and development of educational technologies. Those who attend will become part of a vibrant community of fellows that continues through the year.

Fellowship Benefits

- Full travel and accommodation coverage for summit
- Continuing Education Units (CEUs)
- Direct collaboration with leaders in the field of AI and education
- Year-round professional learning community
- Opportunity to influence development of tools and research

Who Should Apply

- Math teachers
- School administrators
- District math leads
- Math coaches

Key Information

- Summit Dates: June 23-24, 2025
- Location: [institution blinded for review]
- Cost: Fully funded
- Format: Summit plus ongoing collaboration
- Audio and video recording will take place for research purposes

**Apply to the Fellowship**

Submit your application by March 30. We are looking for fellows with all different perspectives to help shape the future of language technologies. We will not have capacity for all applicants and will confirm invitations after March 30. Questions? [lab email blinded for review]

**Recruitment Email**

Dear [Math Educators],

I'm reaching out from the [lab name blinded for review] to invite you to fully-funded two-day summit at [institution blinded for review] on June 23-24, 2025 called Practitioner Voices: Language Technologies in Math Education.

We're bringing together math teachers, leaders, and coaches with diverse perspectives on AI in education – from skeptics to optimists – to help us reimagine how language technologies like Natural Language Processing can serve mathematics education.

Whether you have reservations about AI in the classroom or see its transformative potential, your perspective is valuable. We need varied viewpoints to help us think critically and dream expansively about what's possible with language technologies in classrooms. Your experiences and insights could help shape not just how these tools work, but what they should (or shouldn't) become.

After a brief application, selected fellows will collaborate with leaders in AI and education. The fellowship includes:

- Full travel and accommodation coverage
- Continuing Education Units (CEUs)
- Opportunities to shape future research directions
- Ongoing connection to our research community

At the summit, we will be hosting workshops and sharing demos and resources related to language technology tools in classrooms. Participants will include teachers, coaches, school leaders as well as a small group of researchers. Those who attend will become part of a vibrant community of fellows that continues through the year.

The short application survey is HERE. Applications are due March 30, 2025. You are welcome to apply with colleagues or others in your network! Feel free to share this opportunity widely. To learn more or apply, contact [lab email blinded for review].

**Recruitment Networks**

Recruitment information was sent out through email, social media posts, and word of mouth to:

- [Network blinded for review]: a portfolio of educational projects funded by [funder blinded for review]
- Mathworlds substack: a newsletter with over 34,000 subscribers focused on Math Education written by Dan Meyer

- Existing Networks at institution [specific groups blinded for review]
- Teachers we had previously worked with

## Appendix B
## Summit Application and Selection Process

Practitioners who were interested in attending the summit were asked to fill out a short application, a breakdown of this applicant pool in Table B1. Applicants were informed within a week of the March 30th deadline if they were invited to participate.

**Application Questions**

1. Role
    a. Teacher
    b. Other
    c. School or District Leader

**For teachers:**

2. Grade and Subjects Taught
3. School
4. District
5. At your school, what percent of students qualify for Free and Reduced Lunch (FRL)?
6. Years of teaching experience

**For School/District Leader or Other:**

2. Job Title
3. School/District/Organization
4. At your school or district, what percent of students qualify for Free and Reduced Lunch (FRL)?

**For all:**

5. Where do you live? (City, State)
6. Please list anyone else from your school, district or network applying with you:
7. What draws you to this opportunity? (a few sentences)
8. What do you hope to get out of the summer summit and fellowship experience? (a few sentences)
9. What is something you are working on developing in your teaching practice or your practice? (a few sentences)
10. Which of the following best describes your current use of artificial intelligence-driven tools in your classroom?
    a. I have never used them and I don't plan to start
    b. I have not used them and do not plan to start this school year - but do plan to start in the future
    c. I have not used them but plan to start this school year
    d. I use them a little
    e. I use them some
    f. I use them a lot

11. You indicated you don’t currently use artificial intelligence-driven instructional tools in your classroom. Why not? Select all that apply.
    a. I haven't explored these tools because I have other priorities that are more important
    b. I don't know how to use these tools
    c. My district hasn't outlined a policy on how to use the technology appropriately
    d. I know something about using these tools, but I'm not sure how to effectively incorporate them into instruction
    e. I'm not sure where or how to find these tools
    f. I have data privacy/security concerns about these tools
    g. I don't think the technology is applicable to my subject matter or grade
    h. I don't understand how artificial intelligence works
    i. I don't believe the technology is appropriate for a K-12 setting because of its potential to be used for cheating
    j. We/I cannot afford these tools
    k. Our tech infrastructure (internet access, devices, etc.) is too old/dysfunctional to support these tools
    l. My district, school, and/or supervisor has banned us of the technology
12. How have you used artificial intelligence in the classroom? Select all that apply.
    a. Exploring new ideas for teaching
    b. Creating lesson plans
    c. Creating material to present to students during class
    d. Creating student assignments
    e. Creating rubrics
    f. Checking for student cheating or plagiarism
    g. Researching a topic I will be teaching to get myself up to speed
    h. Parent/guardian/family communication
    i. Translating speech or writing into English and/or a language other than English
    j. Teacher students how to use AI
    k. Creating exams
    l. Writing recommendation letters for students
    m. Completing paperwork
    n. Grading lower-stakes assignments
    o. Providing feedback to students to help them improve
    p. Communicating with students
    q. Communicating with co-workers and/or supervisors
    r. Tracking student attendance
    s. Grading higher-stakes assignments
13. Please cite specific tools and ways you use them:
14. On the scale of 1-4, how much does the following statement describe you: *I am a skeptic*

*of AI in education*

*15.* What about AI/Language Technology in education are you most excited about?

16. What about AI/Language Technology in education are you most skeptical about?
17. **(If teacher)** In a typical year at your current position, about what percent of your students in your class speak a language other than English at home?
17. **(If School/District Leader or Other)** In a typical year at your current position, about what percent of students in your school or district speak a language other than English at home?
18. What kinds of things have you found helpful in encouraging students to talk to each other about math?
19. What kinds of challenges have you had in getting students to talk to each other about math?
20. What languages do you speak?
21. Are you interested in learning more about opportunities to co-design and partner in research projects with our lab? If so, we'll share more information about upcoming projects.
22. This is a fully-funded conference. We will cover transport and food costs for all, plus accommodation for non-local attendees. (Optionally) Please share any additional financial constraints that you may need assistance with. (We may not be able to cover these things, but we will try.)
23. Any other questions or thoughts you would like to share with our team?

**Table B1**

*Applicant Pool Demographics*

| | | |
|---|---:|---|
| Role | Teachers | 166 |
| | School/District Leaders | 64 |
| | Other (instructional specialist, consultant) | 75 |
| Classroom Exp | Less than 5 years | 27 |
| | 5 to 9 years | 39 |
| | 10 to 14 years | 28 |
| | 15 to 19 years | 31 |
| | 20 to 24 years | 24 |
| | 25 or more years | 17 |
| Self rating on the scale of 1-4: *I am* | 1- Not at all skeptical | 80 |
| | 2 | 141 |
| *a skeptic of AI in education* | 3 | 72 |
| | 4- Very much skeptical | 12 |

**Selection Process**

We tagged all 308 completed application responses to the following questions on a three point scale of High, Medium, and Low based on depth of response and openness to new ideas:

- What draws you to this opportunity? (a few sentences)
- What about AI/Language Technology in education are you most excited about?
- What about AI/Language Technology in education are you most skeptical about?

From the applications with all or all but one High tags, we selected a distribution based on:

1. Their response to the question "On the scale of 1-5, how much does the following statement describe you: *I am a skeptic of AI in education*"
2. Grade Level
3. Years of experience
4. Geography

We also factored in recommendations from trusted sources outside our institution about applicants they previously had great experience with. In a few instances, we also prioritized applicant groups from the same sites who applied together.

**Table B2**

*Participant Demographics*

| | | |
|---|---|---|
| Role | Teachers | 50 |
| | School/District Leaders | 4 |
| | Other (instructional specialist, consultant) | 7 |
| Grade Level | Elementary | 13 |
| | Middle School | 22 |
| | High School | 14 |
| | Other/Across | 11 |
| Classroom Exp | Less than 5 years | 10 |
| | 5 to 9 years | 15 |
| | 10 to 14 years | 10 |
| | 15 or more years | 5 |
| School FRL | <40% | 7 |
| | 40-59% | 6 |
| | 60-79% | 11 |
| | 80-100% | 21 |
| Gender | Female | 43 |
| | Male | 15 |
| | Non-Binary | 1 |
| Race/Ethnicity *Values not mutually exclusive* | Asian | 12 |
| | Black or African American | 7 |
| | Hispanic or Latiné | 14 |
| | Multiracial or Biracial | 2 |
| | Native American, Alaska Native, or Indigenous | 2 |
| | Native Hawaiian or Pacific Islander | 0 |
| | White or Caucasian | 27 |
| Are you a first generation college graduate? | Yes- First generation | 20 |
| | No- Not first generation | 41 |

*Note: These counts do not include demographic data from those who preferred not to answer.*

## Appendix C
## Pre-Summit Codesign and Planning

We hosted two co-design sessions six weeks prior to the summit. Below are the guiding questions and some of the responses our teacher advisory board shared in a brainstorming session.

### Guiding questions:

1. Intros: In your current role, what are some questions that matter to you or that you're spending a lot of time with?
2. What helps you feel **energized** and **empowered** in professional spaces? Like you have agency and a platform to share your ideas?
3. (If there's time) What might help you leave the summit having met your goal from above (Right column)?

**Table C1**

*Responses to goal setting during our co-design sessions.*

| What's one thing you're hoping to learn, feel, or do differently by the end of the summit? |
|---|
| I want to take the ideas back to my classroom and use them with my students to help them connect with the content in more meaningful ways, not just for relevance, but for resonance. |
| I would love to develop connections with other math teachers especially those interested in technology in the classroom. And maybe even come away inspired with new ideas for things to create to improve math education. |
| I want to walk away with a new network of colleagues that I can share ideas with - I want to see ways that i can use the technology in a positive way to support students while at the same time having boundaries so that it's not abused. |
| I hope to feel more confident when it comes to technology use in my classroom. I also worry that my school district is falling behind when it comes to digital resources and their capabilities so I would love to be able to take back a wealth of knowledge to share with the math department so we can all benefit from this new knowledge. When I told my principal I would be attending the summit we talked about how there are a million things the kids can do with their Chromebooks, but they're probably only doing a handful… teachers too. |
| I would like to learn how others are leveraging these technologies to improve their teaching (gamifying math, differentiated learning, extensions), hear stories of how students are using these technologies to improve their learning (elaborate on confusing topics, turn lessons into podcasts, summarizing articles, creating mnemonics ), and what math skills you all think workers will need in the future (ex: how did the advent of the calculator change what math skills engineers needed to know) |

**Appendix D**
**Summit Agenda**

## Monday, June 23

| *Estimated Times* | | | *Session Title* |
|---|---|---|---|
| 8:30 AM | 45 min | | Registration and Breakfast |
| 9:15 AM | 25 min | Whole Group | Opening Session |
| 9:40 AM | 40 min | Whole Group | Grounding in Values and Challenges |
| 10:10 AM | 25 min | Whole Group | Idea Grounding: Math and Language |
| 10:25 AM | 15 min | | *BREAK* |
| 11:00 AM | 60 min | Whole Group | Explore A Tool: AI enters the conversation<br>(Please create a free ChatGPT account before or bring your login credentials) |
| 12:00 PM | 60 min | | *LUNCH*<br>*Buffet in Pre-Conference Center*<br>*Optional: Join a Lunch and Learn Discussion in the Main Room* |
| 1:00 PM | 30 min | Whole Group | Answering some morning questions |
| 1:30 PM | 10 min | | *Transition Time* |
| 1:40 PM | 60 min | Breakout | Breakout Session |
| ~2:40 | 10 min | | *BREAK*<br>*Snacks in Pre-conference Center and 2nd Floor Kitchen area* |
| 2:50 PM | 60 min | Breakout | Breakout Session |
| 3:50 PM | 10 min | | *Transition Time* |
| 4:00 PM-<br>4:30PM | 30 min | Whole Group | Wrap Up Day 1 + Logistics for dinner + **Group Photo** |
| 5:45-7:00<br>PM | | | **Group Dinners** |

## Tuesday, June 24

| | | | | |
|---|---|---|---|---|
| 8:30 AM | 45 min | | *Breakfast and Coffee* | |
| 9:15 AM | 60 min | Whole Group | Making Criteria Actionable | |
| 10:15 AM | 5 min | | *Transition Time* | |
| 10:20 AM | 55 min | Breakout | Breakout Session | |
| 11:15 AM | 5 min | | *Transition Time* | |

| 11:20 AM | 25 min | Whole Group | Showcase from Deep Dives | |
|---|---|---|---|---|
| 11:45 AM | 45 min | | *Lunch Time:*<br>*Buffet in Pre-Conference Center* | |
| 12:30 PM - 1:00 PM | 30 min | Whole Group | Closing session: Where to go from here? | |

## Appendix E
## Breakout Session Summaries

Participants pre-selected their preferred breakout group, where they would spend three hours across the two day summit to dive deeply into one of four application areas: student discourse, teacher talk, curriculum adaptation, or feedback on student work.

**Student Discourse:** This breakout engaged with ways that researchers are analyzing student talk. This group did collaborative work centered on how we can leverage technology to help notice, make sense of, and respond to student math thinking and practices with a particular focus on multilingual learners.

*Central Activity:* Facilitators recorded conversations of participants doing a math task together. These conversations were uploaded as transcripts to [tool blinded for review], an R&D tool developed by our lab. Participants then annotated their conversations in the tool completing a scavenger hunt to look for certain language moves in the conversation. Participants got the chance to compare their noticings with LLM noticings to deeply engage with the opportunities and current limitations of language analysis models on classroom discourse.

*Grounding Tool:* [tool blinded for review], a tool for teacher noticing, annotation, and comparison that hosts classroom transcripts on an interactive website.

**Teacher Talk:** This session focused on scaling high quality classroom observation and teaching feedback through a hands-on exploration of an automated feedback tool and a collaborative design sprint to imagine new possibilities.

*Central Activity:* Participants engaged with the [tool blinded for review] where they could explore the dashboard of measures focused on teacher talk moves during instruction. This group considered the possibilities of this technology for instructional improvement and engaged in expansive thinking to develop ideas for where this technology could go moving forward.

*Grounding Tool:* [tool blinded for review], a platform to which educators can upload classroom transcripts and receive an automated analysis of the talk from their class.

**Curriculum Adaptation:** This group centered on the tension that one-size-fits-all doesn't work well for lesson plans, but designing modifications for the learners in context can be very time consuming. This breakout group investigated how NLP tools can address this challenge.

*Central Activity:* Participants prepared an Illustrative Mathematics (IM) Lesson Plan to teach, both working through the lesson and developing multiple representations to support learners. The group then explored [tool blinded for review] and evaluated how LLM outputs supported the lesson design while considering what this technology would be most helpful for in lesson planning and curriculum implementation.

*Grounding Tool:* [tool blinded for review] an AI interface that supports teachers to adapt IM lessons for their contexts.

**Feedback on Student Work:** This group explored what kind of written feedback best supports student learning in math. Participants jumped into giving and refining feedback as a way to investigate how to assess quality of feedback and how to give more effective feedback.

*Central Activity:* Participants brought in examples of student work and their feedback on the work. They deeply considered and revised this feedback then got to integrate LLM generated feedback. This generative comparison brought a deeper understanding of the potential and limitations of this technology.

## Appendix E
## Grounding Slides for Summit

# Why are we here?

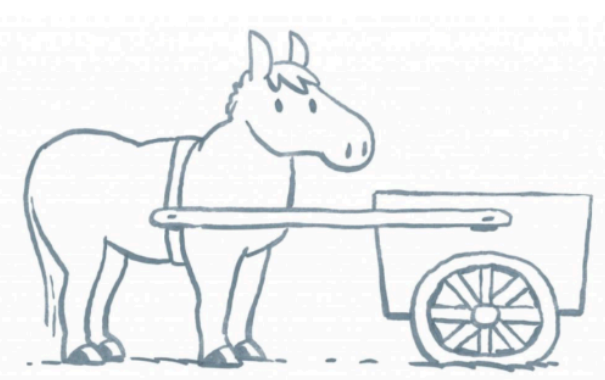

Often, **educators are given tools** that have been developed and give feedback post hoc

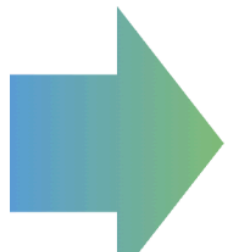

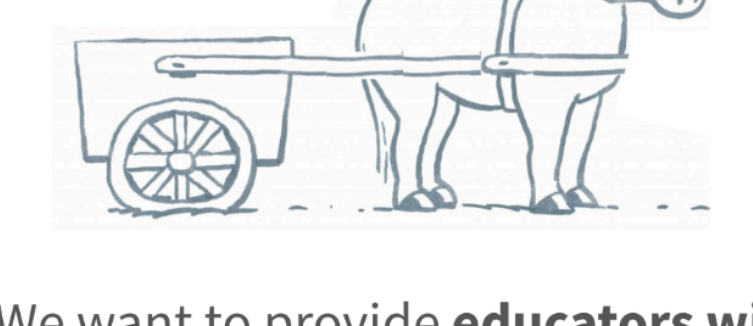

We want to provide **educators with a platform to create the vision and direction** for ed tech research and development (R&D)

# Shared Perspective: Curious, Critical, Hopeful

We move in and out of these perspectives. You may have more experience with some of these approaches to GenAI and Language Technologies in Math Education.

Challenge yourself to **strengthen the muscles** you may have less practice with.

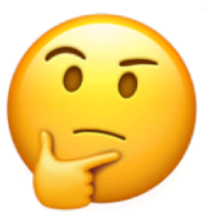

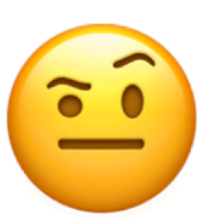

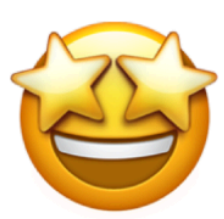

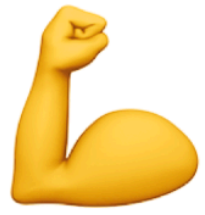

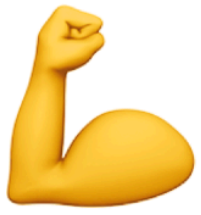

## This summit is a platform for…

- Teacher voices and experiences
- Discovery and sharing
- A variety of expertise
- Networks, connections and ideas

to inform research and development on teacher-facing edtech.

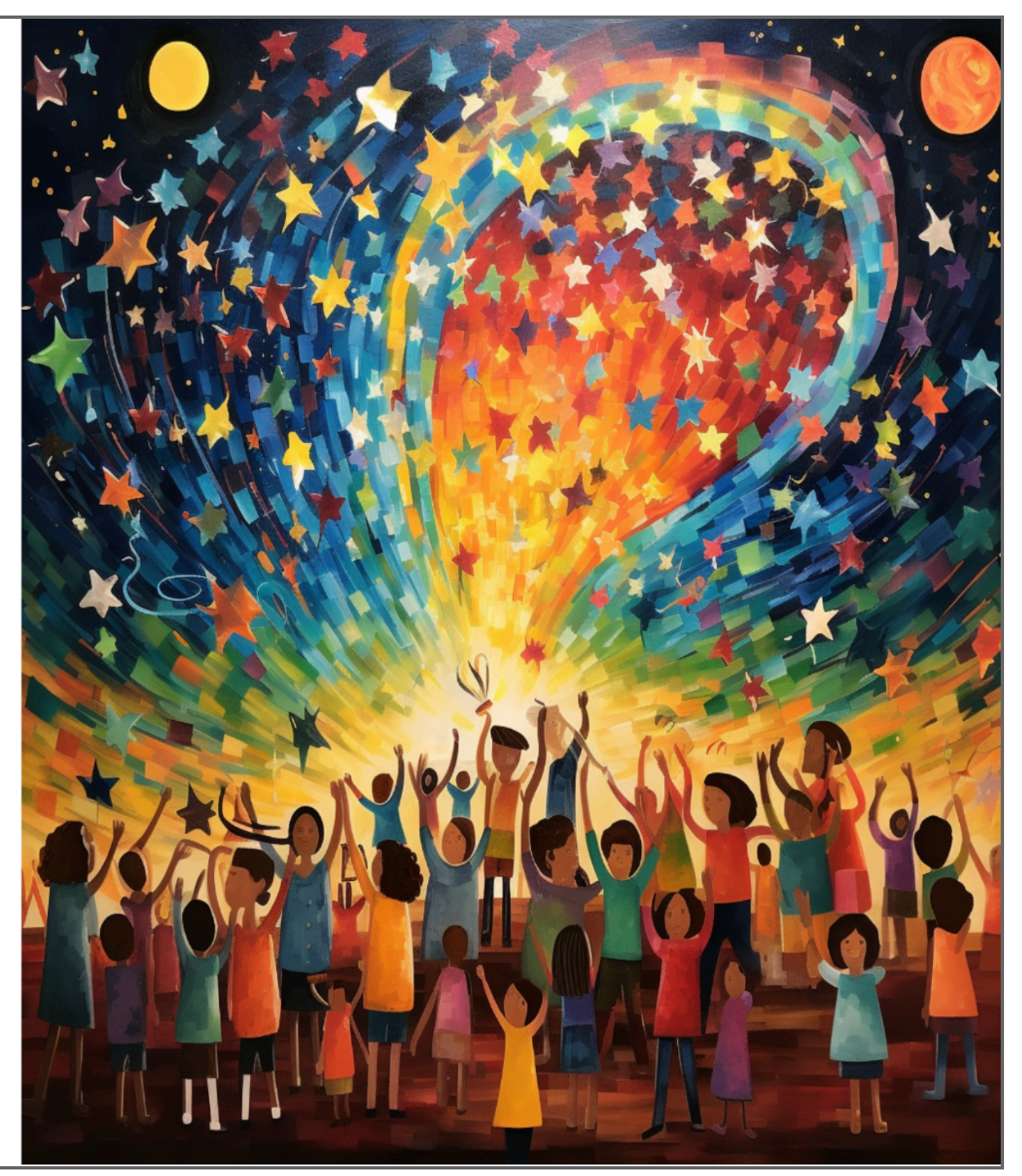

## This summit is NOT …

- Training on specific AI tools
- Focused on resources to use with students
- One-sided extraction of information for research
- An EdTech expo
- Taking a position on AI use by students

shifting the focus on the **learning process** and on **curiosity** and **motivation**

## Appendix G
## Sample of Group Criteria Charts from Explore a Tool Session

During the Explore a Tool Discussions, in which table groups determined a scenario and used Chat GPT to address their situation, we prompted groups to collect their ideas on anchor charts with a list of important criteria to consider with reasoning. We displayed these charts for for the rest of the summit for participants to engage with.

*Sample selection of Group Anchor Charts*

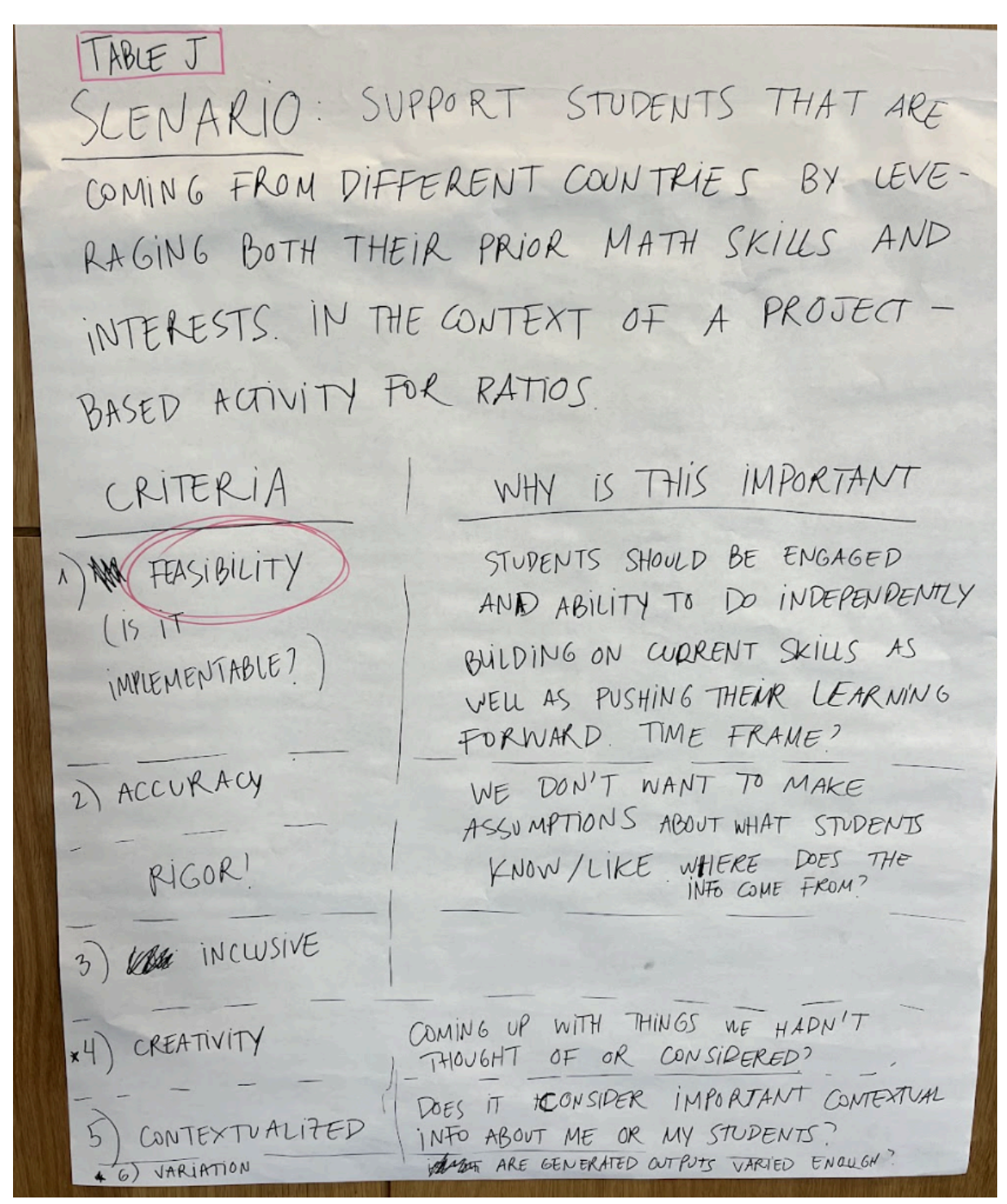

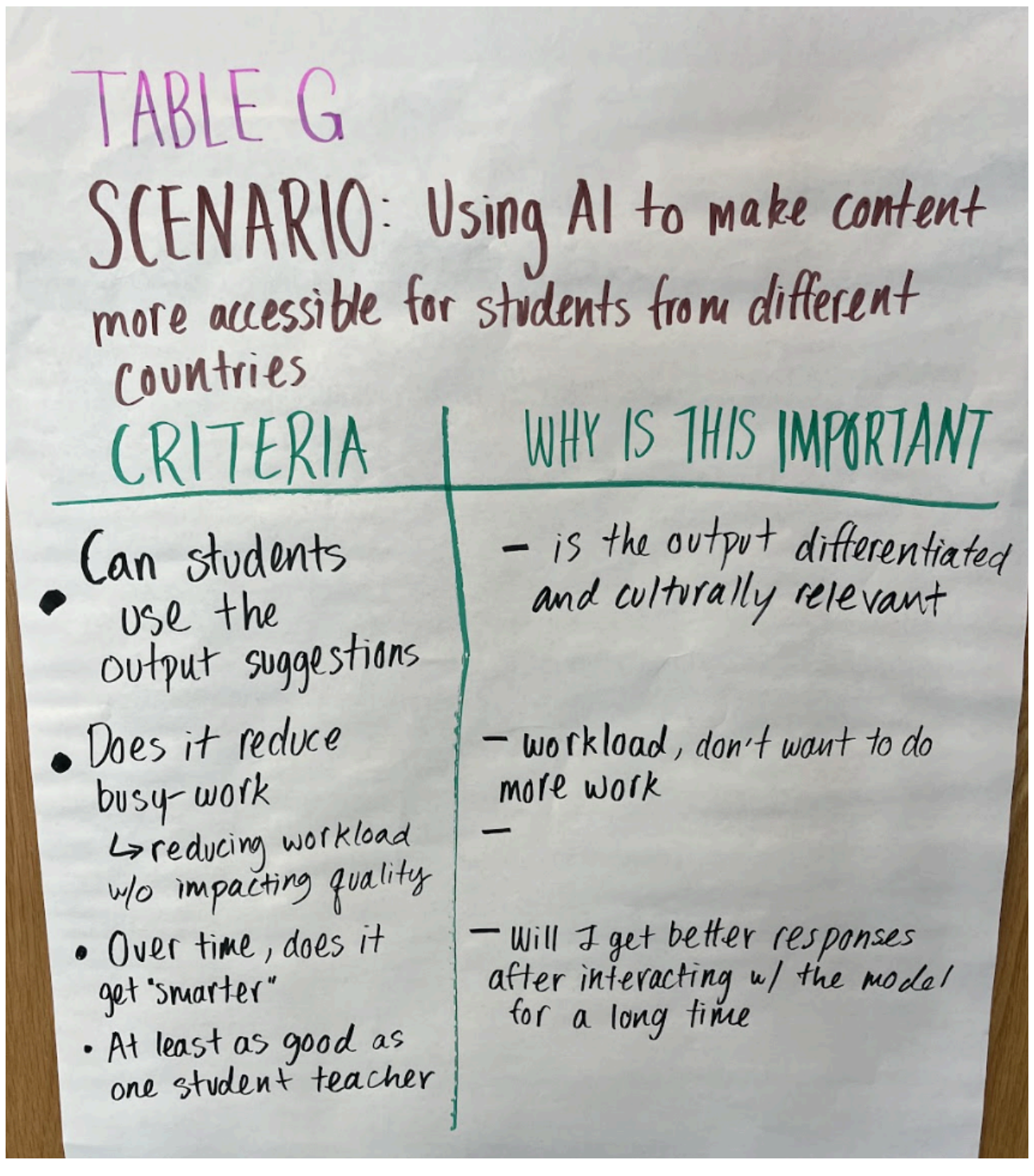

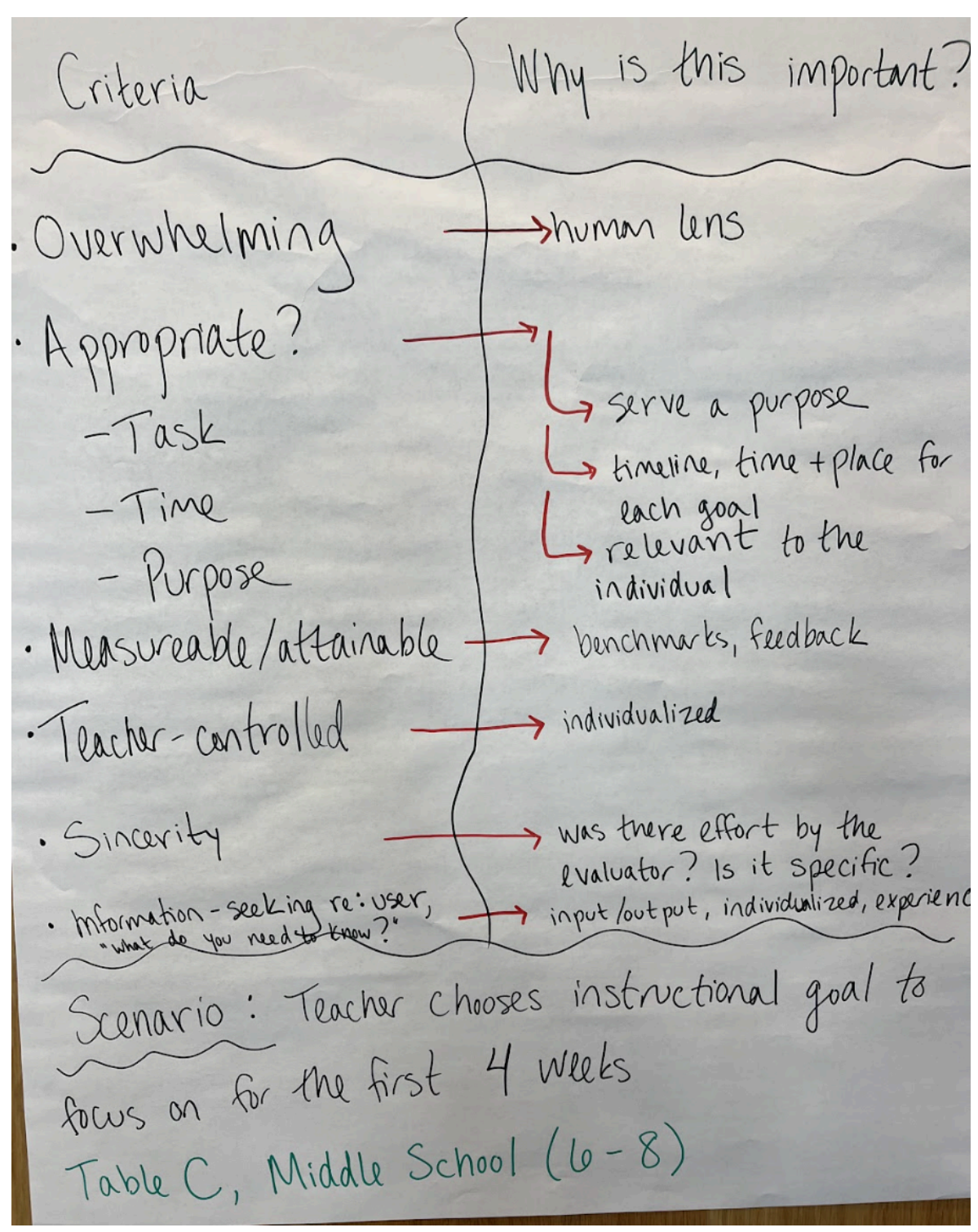

**Appendix H**
**Instructions for Rubric Activity**

We began day two by reflecting back some patterns we found in their individual criteria selections from the day before. This grounded rubric conversations not only in individual criteria, but also in the trends shared across the group.

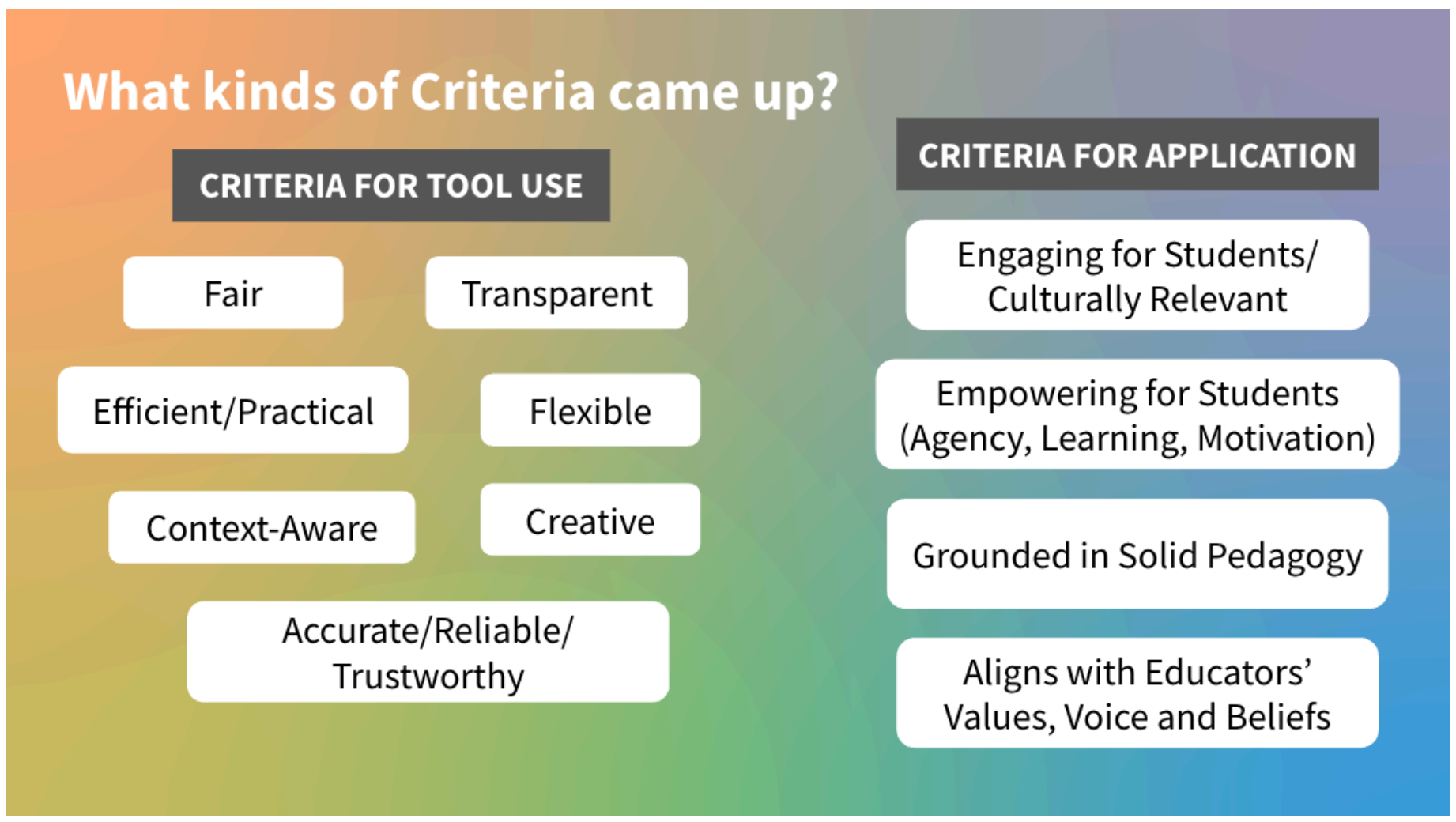

We contextualized the activity with the following prompts:

**Zoom Out: why are we doing this?**

- We want your thoughts to guide the people building things for classrooms (who don't know much about classrooms)
- Give you a chance to represent your students and peers and give voice to what matters to them
- Support you in building a tool to support you to have productive and grounded conversations moving forward

**What are we going to do with these?**

- Look at them all together to share the story of what matters to you. (will be aggregated, so wording doesn't need to be perfect)
- Your input will be one instrument in a symphony we share widely

We provided the following instructions for the rubric development activity:

**Build a personal rubric based on the criteria that matter to you:**

- Don't need to shy away from **personal values**- what matters to you should show up in the rubric
- Consider the **PURPOSE** of your rubric and the **AUDIENCE**

- Frame and describe things for someone who is less well versed in classroom contexts

**Things to consider during your rubric work time:**

- **List specific criteria:** Can you notice it? Does it need to be broken up even more?
- Share **descriptions of low and high** achieving traits (don't need to get lost in the weeds, don't need to build out the difference between 2 and 3 on the scale)

Participants then had time to share with a partner, ask questions of each other's work, and revise or add to their own rubric draft before sharing out in their small groups for 15 minutes.

# Appendix I
## Sample of Personal Rubrics

Audience: School Staff

Purpose: To assess how well an AI tool supports student-centered learning in a middle school setting

| Criteria Description | Low Level | High Level |
|---|---|---|
| Enhancing Learning | Students are passive users; the tool adds little to their understanding or curiosity. Most interaction involves copying and pasting with minimal original thought. | Students actively engage, think critically, and deepen understanding. They originate their own ideas, build on concepts, and use the tool to support meaningful, creative thinking. |
| Adaptability | The tool doesn't adjust to students' different needs or learning goals. | Students can use the tool in ways that fit their learning styles, interests, and pace. |
| Low Floor / High Ceiling | Students struggle to begin using the tool or quickly run out of ways to grow with it. | Students of all skill levels can jump in easily and explore deeper challenges as they grow. |
| All Students / All Staff | Only some students can access or benefit from the tool; others are left out. | All students feel included and supported—tool works for a wide range of learners and teachers. |

Audience: Teachers, PLC community

Purpose : Tool for lesson planning + Student Assessment

| Criteria | High | Low |
|---|---|---|
| Accurate | - Content is factual & supported by research.<br>- Provides research based methods. | - content is inconclusive lacks consistency or is not scientific in nature |
| Context - Aware | ⊙ Inclusive of various levels, age groups & backgrounds of learners<br>⊙ Relevant & applicable suggestions | - narrowly addresses or caters to different audiences<br>- Content is not applicable relevant or realistic. |
| Culturally Relevant | - Diverse examples<br>- Covers a wide cross Section of cultures & experiences<br>- use of appropriate language | - Limited background + narrow audience<br>- Limited language use |
| Efficient | Provide a wide cross<br>- Section of information<br>- includes choices that are Practical<br>- Room to Explore search fine tune or Expand | - Limited knowledge Scope & does not Provide extension of Content |

Rubric Audience: Ed Tech

Rubric Purpose: Is this platform asking students to engage in complex mathematical thinking?

| Criteria Description | Low Level | High Level |
| --- | --- | --- |
| Is there only one correct answer to questions? | Yes | No |
| Are there multiple entry points for students to engage with the task/question? | No | Yes |
| Does this platform allow for students to explain their thinking in multiple modalities (voice, text, drawings)? | No | Yes |
| Are the tasks asking students to apply mathematical understanding to real world scenarios (that are actually reasonable, stereotype free, and culturally relevant)? | No | Yes |
| Does this platform allow for meaningful collaboration between students and groups of students? | No | Yes |

## Appendix J
## Registration and Pre-Summit Survey

Questions related to logistics are excluded from the list.

1. Please rank your order of preference for the following breakout rooms:
    - Feedback on Student Work
    - Curriculum Adaptation
    - Teacher Learning
    - Student Discourse
2. (optional) Anything you want to share about your decision?
3. Which grade level of band would you like to be in small groups with? (If you are a teacher, please pick your own grade level band)
    - Lower Elementary
    - Upper Elementary
    - Middle School
    - High School
4. What’s one thing you’re hoping to learn from the summit?
5. What’s one thing you’re hoping to share at the summit?
6. What kind of impact do you think AI will have on math instruction?
7. (optional) Do you have any relevant experience, specific training or expertise in instruction, technology or teaching that you would like us to be aware of?
8. Any other questions, comments, thoughts you'd like to share?

## Appendix K
## Post-Summit Survey

### General Summit Feedback

1. Overall, how productive do you think the summit was for you?
2. How likely would you be to recommend this event to a friend or colleague?
3. Did you feel that you had enough information to prepare for the summit?
4. (optional) What other information or communication would have been useful to you?
5. Did you feel that you were financially supported enough to join this summit?
6. (optional) Any other thoughts to share on financial support?
7. What were some of the most valuable components of this summit?
8. Which summit activities best scaffolded your reflection on AI tools in your practice?
9. What other activities might scaffold further reflection?
10. What topics or activities would you like to see expanded or added in future events?
11. If you could tell people working on developing these technologies one thing to keep in mind, what would it be?
12. What kind of impact do you think AI will have on math instruction?
13. What's something you have learned or are taking with you from the summit related to your values and challenge from Day 1? (Final Turn and Talk question!)
14. (optional) Are there any moments from the summit that are especially memorable that you'd like to capture?

### Summit Goals & Sessions

15. How well do you think the summit supported you to reach the following goals:
    - Discover: engage with NLP and LLM tools and discover their capabilities and limitations related to math instruction.
    - Share: share insights and dream expansively around how language technologies could address goals and challenges they have or cause harm.
    - Energize: leave with new connections, new ideas, new questions, and next steps.
16. How did each of these sessions support your learning and connection:
    - Value and Challenge Reflection (Re-Ignite Cards)
    - Explore a tool (Developing Criteria)
    - Developing a Personal Rubric
17. How effective were whole group facilitator(s) in creating a clear, inclusive, and engaging session? (Includes a follow-up "Why did you choose that rank?")
18. How do you feel about the quantity of content we covered in Whole Group Sessions?
19. (optional) What thoughts do you have about the quantity of content covered in Whole Group?

### Breakout Session Feedback

20. Which breakout session were you in?
21. How effective were your breakout facilitator(s) in creating a clear, inclusive, and engaging session? (Includes a follow-up "Why did you choose that rank?")
22. What's one thing your facilitator(s) did that supported your learning or engagement today?
23. And what's one thing they could do differently next time to improve the experience?
24. What did you appreciate about the session facilitation and activities?
25. What insights did you gain from the session that you want to share with us?

**AI for Student Feedback**

26. How likely would you be to use an AI tool for providing feedback to students, if it were available to you? (Includes a follow-up "Why did you choose that rank?")
27. In what context would you feel most comfortable and find the most value using such a tool?
28. What do you see as the single biggest potential barrier that would prevent you from using such a tool in your practice?
29. What is the most important thing a school or district would need to do or say to earn your trust in using a tool for feedback provision?
30. Of the different ways that you "taught" AI about good feedback (prompting, providing your own examples, ranking AI examples), which of these do you feel is the best / most effective use of your expertise?

**AI for Curriculum Enactment (e.g., Coteach.ai)**

31. How likely would you be to use an AI tool for curriculum enactment, like Coteach.ai? (Includes a follow-up "Why did you choose that rank?")
32. What support would you want to help you use such a tool?
33. What do you see as the single biggest potential barrier that would prevent you from using such a tool in your practice?
34. What is the most important thing a school or district would need to do or say to earn your trust in using a tool for curriculum enactment?

**AI for Student/Classroom Talk (e.g., M-Powering Teachers)**

35. How likely would you be to use an AI tool that analyzes student talk / classroom discourse (e.g., M-Powering Teachers), if it were available to you? (Includes a follow-up "Why did you choose that rank?")
36. In what context would you feel most comfortable and find the most value using such a tool?
37. What do you see as the single biggest potential barrier that would prevent you from using such a tool in your practice?

38. What is the most important thing a school or district would need to do or say to earn your trust in using a tool that analyzes your classroom conversations?

**Future Engagement**

39. Through what channel(s) would you be interested in staying in touch?
40. Would you be interested in serving as an ambassador/facilitator for this community network? (e.g. hosting a virtual or in-person meetup, moderating the Slack channel, sharing resources with the community)
41. Tell us more about what you would like to take on or how you would like to contribute.
42. Would you be interested in participating in a (max. 30 minute paid) follow-up interview?
43. Are you interested in attending similar events in the future?
44. Would you be interested in supporting the research (e.g. data analysis, writing) being conducted from this summit?
45. (optional) Is there anything else you'd like us to know?

## Appendix L
## Post Summit Interview Recruitment and Protocol

**Recruitment Email**

Hello ___!

This is a late, great follow-up from the Practitioner Voices Summit this summer. We've been going through the conversation data from the event, and we would be interested in interviewing you to better understand your experience. The goal of this series of interviews is to explore how connecting with other educators influenced how you think about and evaluate AI.

This would be a 30 minute Zoom call with the research team that gets recorded. As an appreciation, we would send you a $30 giftcard.

If you are interested in participating, please respond with the blocks of time you are typically available (ex. Tues, Wed, Thursday after 3:30 pm CT, 10:45-11:30 am CT every weekday). The holidays are upon us, but we'd like to schedule each of these interviews before Dec 12th.

Thank you for all of the work you do to support deep learning,
[The research team]

**Interview Protocol**

1. ["We can stop at any point, if you don't understand a question let us know and we can rephrase…"] We have survey responses, so we don't need an overall evaluation from you. We're getting at what was sticky in this interview.
2. We know it's a long way back, but what conversations from the summit do you still remember and think about?
3. We'd like to play you a clip/show you a transcript from a conversation you were a part of. Focus on your specific role:
    a. How do you think others influenced your evaluation criteria during this conversation?
    b. How did you intend to influence others' evaluation of AI tools during this conversation?
4. Imagine that you had gone through the steps of the summit by yourself (reflecting on values, coming up with a challenge, exploring how a tool might help you address that challenge, creating an evaluative structure for similar tools). How do you think that experience would have been different from the experience you had?
5. (If time allows) Have you ever participated in AI-focused PD or research before?
    a. What were those experiences like?

## Appendix M
## Heterogeneity Analyses

### Methods

In addition to analyzing overall criteria distributions, we explored potential subgroup patterns by cross-tabulating the frequency of codes and themes with teacher metadata collected during registration. We considered three participant-level variables: grade band/role (Elementary, Middle, High school, Other/Leader), an indicator for high free/reduced-price lunch (≥80%), and an indicator for 10+ years of experience. We estimated a pooled multinomial logistic regression at the criterion level ($n=297$) to assess whether teacher demographics predict which category of criteria teachers emphasized (Equitable, Flexible, Practical, Rigorous). We used cluster-robust standard errors at the teacher level (55 teachers) to account for repeated observations per teacher. Rather than interpreting coefficients relative to an arbitrary reference outcome, we summarized results using average marginal effects (AMEs), which quantify how each demographic predictor changes the predicted probability of each category in percentage-point terms. For grade band, a categorical predictor, AMEs are reported relative to the omitted reference grade band (Elementary). Given the modest sample size, these comparisons are treated as exploratory, using descriptive patterns to illustrate trends rather than to make systematic claims about population-level differences.

### Results

Our exploratory heterogeneity analysis examined whether teachers' demographics predicted which category of criteria (Equitable, Flexible, Practical, Rigorous) they most emphasized. As expected due to the modest sample size, we observed few reliable subgroup differences overall, and we treat these patterns as descriptive rather than definitive. By grade band/role, High school participants were 12 percentage points less likely to emphasize Equitable criteria than the Elementary reference group ($\text{AME} = -0.121$, $p = 0.030$; 95% CI [−0.230, −0.011]), with an offsetting trend toward greater emphasis on Practical criteria ($\text{AME} = +0.141$, $p = 0.080$). The merged Other/Leader group was 13 percentage points less likely to emphasize Flexible criteria ($\text{AME} = -0.135$, $p = 0.001$; 95% CI [−0.217, −0.054]); corresponding increases in other categories were positive but not statistically significant. Middle school participants did not differ from Elementary for any category (all $p$'s $> .40$). For other demographics, effects were limited: high FRL (≥80%) showed a marginal decrease in the probability of emphasizing Rigorous criteria ($\text{AME} = -0.058$, $p = 0.089$), but did not predict Equitable, Flexible, or Practical emphasis (all $p$'s $\geq .20$). Having 10+ years of experience was not associated with shifts in any category (all $p$'s $\geq .41$). Overall, the clearest patterns were lower emphasis on Equitable criteria among High school teachers and lower emphasis on Flexible criteria among Other/Leader participants, relative to the Elementary reference group.

# Appendix N

## Table N1

*Codebook for Surveys and Interviews*

| Parent Code | Sub-code | Definition | Example (from Survey) |
| --- | --- | --- | --- |
| Structural | Breakout Sessions | References to small group breakout sessions as valuable summit components. Includes mentions of specific breakout topics (e.g., Teacher Talk, Curriculum, Feedback) and the focused nature of smaller group work. | "The breakout room (feedback) was very informative and helpful for giving students feedback in math!" |
| Structural | Time for Discussion | Emphasizes the importance of having adequate time for conversation, dialogue, and collaborative discussion. Includes both appreciation for discussion time provided and requests for more time. | "It was so valuable to interact with teachers from around the country to find commonalities and see differences that we experience in the classroom." |
| Structural | Hands-on Activities | References to interactive, experiential learning activities where participants actively engaged with AI tools or materials rather than passively receiving information. | "Having hands on activities in groups and open conversations with all people participating in the summit." |
| Structural | Diverse Groupings | Highlights the intentional mixing of participants from different backgrounds and experiences with AI in group configurations. | "Being exposed to people in that group who had limited experience with AI really motivated me to think about how do I communicate, and how do I support teachers who have limited, experience with AI." |
| Structural | Organization | Comments on the overall structure, logistics, flow, and planning of the summit. Includes appreciation for clear agendas, smooth transitions, and well-coordinated activities. | "The organization of the summit!" |
| Content | AI Understanding | Deepened comprehension of how AI/LLMs work, their capabilities, limitations, and underlying mechanisms. Includes technical knowledge gained about AI tools. | "Getting a full understanding of how Large Language Models work...see not just the successes of AI but also the failures." |
| Content | Tool Exploration | Hands-on experience testing and experimenting with specific AI tools (ChatGPT, Claude, CoTeach, etc.). Includes comparing tools and discovering their practical applications. | "The activity where we needed to tackle a challenge as a group using ChatGPT and comparing and contrasting our results and how we structured our prompts." |
| Content | Grounding | Connecting AI use to personal values, teaching beliefs, and professional identity. Includes reflection on how values (e.g. generosity, creativity, joy, justice, health/wellness, independent, equity, environment) guide technology decisions and classroom practice. | "It serves as the foundation that helped me ground my evaluation and expectation of AI in my specific teaching context." |
| Content | Rubric/Criteria Development | Creating evaluation frameworks, criteria, or rubrics for assessing AI tools. Includes articulating standards for what makes an AI tool effective or appropriate. | "Writing the rubric to evaluate AI. It helped to clarify what I want AI to do for me." |
| Content | Classroom Applications | Specific ideas for implementing AI in teaching practice. Includes concrete use cases, lesson ideas, and practical strategies for integrating AI with students. | "A better cohesive view of what AI is behind the scenes, and several new ideas from other teachers of how to incorporate technology into helping students feel seen and heard." |
| Interpersonal | Networking | Building professional connections and relationships with other educators, researchers, and leaders. Includes exchanging contact information and establishing ongoing relationships. | "The networking and relationships built were invaluable components of the summit! I can't put a price on that!" |
| Interpersonal | Collaboration | Working together with others on shared tasks or goals. Emphasizes joint problem-solving, co-creation, and teamwork during summit activities. | "The opportunity to collaborate with other teachers." |
| Interpersonal | Diverse Perspectives | Exposure to and appreciation for viewpoints from people with different backgrounds, experiences, roles, or opinions about AI in education. | "A lot of people with diverse perspective with whom I can disagree really helped build consensus and a better understanding of our own practices." |
| Interpersonal | Gained Confidence, Inspiration, Motivation | Increased self-efficacy, intentionality or comfort level with AI tools as a result of summit participation. Includes feeling more prepared/motivated or less intimidated to try AI. (This code especially relates to agency.) | "I am leaving feeling more confident, understanding from different perspectives and new tools/strategies." |
| Interpersonal | Feeling Valued | Sense that one's voice, expertise, and contributions were respected and appreciated. Includes feeling listened to and having input taken seriously. | "Our ideas and voices were truly well-listened to and valued there." |

## Appendix O

**Table O1**

*Code Totals for Research Question 2*

| Parent Code | Sub-code | Survey Count | Interview Count |
|---|---|---|---|
| **Structural** | Breakout Sessions | 27 | 3 |
| | Time for Discussion | 28 | 2 |
| | Hands-on Activities | 12 | 5 |
| | Diverse Groupings | 3 | 5 |
| | Organization | 9 | 2 |
| **Content** | AI Understanding | 23 | 5 |
| | Tool Exploration | 23 | 5 |
| | Grounding | 19 | 4 |
| | Rubric/Criteria Development | 26 | 4 |
| | Classroom Applications | 22 | 6 |
| **Interpersonal** | Networking | 16 | 1 |
| | Collaboration | 12 | 5 |
| | Diverse Perspectives | 25 | 6 |
| | Gained Confidence | 10 | 5 |
| | Feeling Valued | 8 | 1 |

## Appendix P
## Table Discussion Excerpts Illustrating Collaborative Reflection

The following two excerpts are drawn from recorded table discussions during the whole group ChatGPT exploration activity. In this activity, small groups of teachers shared a classroom challenge, explored how ChatGPT might help them address it, and generated evaluation criteria based on their experience. Excerpts are presented unabridged. Bolding reflects emphasis added by the authors to highlight key moments.

### Excerpt 1: Productive Pushback (Table I)

In this excerpt, teachers debate the utility of ChatGPT for generating culturally responsive word problems. The discussion is driven primarily by John, who takes a skeptical stance, arguing that ChatGPT is not useful because it fabricates data rather than drawing on real sources and takes too much effort to produce quality output. Several peers push back — both making specific suggestions for better prompts and arguing that, like any tool, AI requires practice and learning over time. Notably, some of those pushing back share John's reservations about AI, yet still press him to distinguish between principled critique and blanket rejection. Over the course of the conversation, John's criteria become more precise: he moves from a general complaint about fabricated data to specific expectations about default sourcing and transparency. The group also surfaces a deeper tension between the time investment required to use AI well and the efficiency gains it promises.

| Line | Speaker | Text |
| --- | --- | --- |
| 1 | John | I looked at a prompt related to social justice math in Mexico and [ChatGPT] gave me a basic scenario... I said, can you make it a little bit more complicated? Then I even tried to do something with reparations and it gave me two options or scenarios but I want actual data. There's no sources here. Where's the data coming from? I'll just give you an example, right? In the problem scenario about Mexico, using systems of equations, it talks about fair wages. And then it says, you know, Anna worked 40 hours and earned blank pesos. Jorge worked 35 hours and earned blank pesos. You overhear that Jorge may be earning more for hours because he's male, you want to figure out the hourly wage like... It's all made up. |
| 2 | Pamela | But most math problems are made up. [laughter] |
| 3 | John | No, no, but there is actual data that you can access this information from about males and females working in these countries. And so I think that... |
| 4 | Sarah | Did you ask it like, what is the actual data, where did you get your statistics from? I know sometimes AI lies to you straight up... |
| 5 | Pamela | Actually, so, I was working with a math provider for the district that had to create "relevant" [air quotes] tasks. So I actually did run with fair wages, and I got data from the wage gap. |
| 6 | John | [interrupting, pointing at Sarah] So just to add to it — it said, great question, the specific hours, wages and names used in the scenario created are fictional, but "plausible" [air quotes], crafted to reflect real world patterns. |
| 7 | Pamela | Yeah. |
| 8 | John | So it's not — they — |
| 9 | Miguel | Yeah. |
| 10 | Sarah | Like at that point, that's not a social justice question. See, that's just a math problem. |
| 11 | John | Right, exactly. And that's where, yes it's putting it in this wage gap thing, but it's basically a word problem that is made up for this scenario. And I would want my students to... Now see. It does... So, I admit, when you led me, [Sarah], to ask that question, it did give me a little bit more information, right? It has data-informed inspiration, but not direct resources. It gives me OECD, gender wage gap data, Mexican government labor ministry, things like that. I would hope that I wouldn't have to ask multiple times for that. Like that should come up right away, to be able to help... |
| 12 | Pamela | [laughs] |
| 13 | Sarah | Not that I'm defending AI though, but that is how it would work with a human though, right? I still have sticky notes of like math problems that my students made me at the end of the school year. But I just said, "Hey, can you make me some FOILing problems?" And they gave me FOILing problems, but then I had to sit down and say, "Hey dude, where did you find those numbers? How did you know that was the answer?" You know, like — not that I'm the biggest proponent of AI, because it lies straight to your face with a smile — but you know, to me that kind of... I want it to have the data straight up front, but like nobody gives you the data straight up front. You have to ask. |
| 14 | John | But, if you're trying to save time — |
| 15 | Pamela | Start with the data. |
| 16 | John | They... |
| 17 | Pamela | I say start with the data. If you're really trying to save time, start with the data set and say, "Use this data set to create problems." And then I think you'll get better results. |
| 18 | Alex | I was, I was gonna — actually, I wasn't thinking exactly that, but I was thinking along the lines of — some of that can help with like the prompts that you put, |

because something you could do is in your initial prompt, like you could say at the end, any data that you use, you must include why you think this is reasonable.

| | | |
|---|---|---|
| 19 | Pamela | Yes. |
| 20 | Sarah | What I wonder is — and I'm not doing this cause I'm lazy — but I wonder if, even if you came to it with the prompt of, using actual data from the Mexican Labor Government, could you make me word problems for a systems of equations? |
| 21 | John | So at some point, if we're spending so much time trying to perfect the prompt |
| 22 | Sarah | Oh yeah, yeah. |
| 23 | John | At what point is it easier for you to just google it? |
| 24 | Sarah | Do it yourself. |
| 25 | Pamela | But I will say this, because I pay for ChatGPT — I pay 20 dollars a month — it's worth every penny. It's just like any other tool, you get in, the more you use it, the better. So it's not going to save you a lot of time up front, when you are first learning how to use it. But it knows me pretty well now, and so it gets better over time. I've been using it, pretty much — I've paid for it for about a year — and it gives me some really good stuff. But it's because I've learned what prompts to give it. But when I first started, I was not using it very well, and it was a love-hate relationship. [laughs] |
| 26 | Sarah | My prompts are — just because I've used it before — my prompts are pretty long. You know, so I kind of go at it with, I want exactly this, I want to emphasize this, I want to specifically include this, I want to do this. Like, to your point [pointing at John], it is a lot of work up front, but I feel like that kind of helps. Like it's learning from you, but you're also learning from it how to interact. |
| 27 | Alex | Yeah. Like we would never in a million years tell somebody, like, "Oh my gosh, the very first Word document you ever made looks like crap, there's no headers, there's no whatever." You know. And there's a bunch of frameworks out there. I did like a Google AI for Educators little thing, it takes like an hour, but they sort of have frameworks... So anyway, all this to say, like there's this weird duality, cause like what we're talking about — our voices ought to be included in the design of the thing. Like, why isn't there a pedagogy-optimized chatbot? |
| 28 | Sarah | Yeah, absolutely. Like, why isn't there — like, it's almost like, you know, if you want to tie it to a specific real-world thing, like that kind of needs to be its own thing, so like it automatically knows to do that. |
| 29 | Judit | So, can I say this — this is where my question came from. There have been curricula designed that, for example, are socially relevant, social justice, they're related to the real world. They've been around for a long time. So they may be out of date. But if they're not accessible to ChatGPT, then it's not having access to the pedagogically designed curricula that have been piloted with real kids, all over the world. And for me to not know that — I mean, we should all know, or just figure it out. But they're probably not, because they are owned by people and they're PDFs. When I see teachers working this hard to create curriculum, I wonder, what did we create all those curricula for, that had all that wonderful pedagogy? Am I making sense here? It's too much work to teach ChatGPT, when you all know how to look at a curriculum and say, "Oh, that one is pedagogically sound." |
| 30 | John | Yeah. |
| 31 | Judit | Which is why I tell teachers: start off with the curriculum that is already mathematically sound, and adapt it to your kids. |
| 32 | Pamela | I mean, that's what I tell teachers. Start off with — there's all this great work that people have already done. It's already got the answers. You know that it's right? Start with that. And then I ask ChatGPT to adapt it to the kids that are sitting in my class. And I tell them what they're interested in, because I know them, because I took the time to find out what they are interested in. And that's what I put into ChatGPT. |

**Excerpt 2: Idea-Building (Table E)**

In this excerpt, teachers have just spent several minutes exploring ChatGPT independently and now share the prompts they used and the outputs they received. In contrast to Excerpt 1, where peers pushed back against a skeptic's stance, here teachers build on each other's ideas. The conversation surfaces two general prompting approaches: a prescriptive approach that requests specific lesson materials (illustrated by Mark's prompt) and an open-ended approach that targets broader pedagogical goals (illustrated by Andres and Paule's prompts). Andres explicitly names this distinction, observing that prompt specificity shapes whether AI functions as a lesson-writing assistant or a broader brainstorming partner. Paule extends this reasoning to instructional coaching, describing how she prompted ChatGPT to help her coach teachers rather than simply produce materials. Later, Nikkie introduces the strategy of assigning AI a role, which other teachers take up and push in new directions — imagining AI as a thought partner that can simulate student thinking, administrator perspectives, or professional communication. Together, the group expands the space of possible uses for AI while simultaneously sharpening criteria for quality: fit to purpose, pedagogical leverage, and the teacher's ongoing role in steering outputs.

| Line | Speaker | Text |
|---|---|---|
| 1 | Mark | My initial prompt (thank you for asking) was, "Create the best lesson for 7 RPA 1 that will integrate joy, humor, and fun." So just kind of selecting a specific standard. And then, seeing what pops out. So, it's a pizza party. You're creating a menu and pizza. Tells you the materials you need. Writes some objectives for you. There's a hook. A little mini lesson with like, you know, reviewable things. And then the actual activity. And a share out and an exit ticket. Shazam. |
| 2 | Andres | So yeah, so yours was very like, "Give me materials." When I did mine, I was like, "I want to make my class more fun and engaging for students to help motivate them." And it went into giving me different prompts. It was like, "Oh, you can gamify learning with different types of games, making maybe a point system, making it more engaging through hands-on movement-based learning, creating stations, gallery walks, scavenger hunts, giving student voice, menu boards, classroom jobs, let them lead, classroom environment, giving different seating." So, I thought that was cool, like different things that you can try in your classroom to really go through. Different routines, and then it's like, build routines that include check-ins, high-low of the day. So it was like we just gave you different prompts to kind of go through. |
| 3 | Carwai | My prompt was like, "Design a lesson on linear functions for eighth grade algebra. I want the lesson to give students the understanding of what our topic is, our strength, focus. The lesson needs to be after the initial introduction. I want to be able to provide positive support to students and help them refine their understanding." |
| 4 | Mark | Is that all? [laughter] |
| 5 | Carwai | It sounds a lot like what yours came out like [inaudible]. The only thing for me that stands out is it starts out by saying, "Think about something you understood well from our first lesson, write it down, and then what made you pause or unsure? What might be a strength in disguise?" That's the only thing that really [inaudible] positive piece. Everything else is like a lesson plan, process, materials. |
| 6 | Paule | I love what mine gave me. I'm sitting here having a little conversation and I'm about it. So the first prompt that I asked ChatGPT: "How do you utilize AI to incorporate belonging in a math classroom where it is crucial to deeply know each individual student?" It basically... this was a lot of things that I already know. Talking about differentiation. AI-assisted insight for deeper relationships. So tools like AI dashboards, to what you guys are talking about with EdLight. Who's taken the risk lately, who thrives on peer interaction, who shuts down during word problems, creating math prompts with AI co-pilots, and then with that belonging piece, getting student feedback. So, *then* I asked it (or I told it), "I'm an instructional coach for a middle school. How do I coach a teacher on how to incorporate what you're saying?" Then this stuff is so juicy. So, sorry if I'm taking up too much air time, but I'm really excited. You can tell me to stop if you think it is too much. Framing... so the coach move: invite reflection with the question, "What does it look like for every student in your class to feel confident, included, and seen in math?" And then it gives me a goal. Starting small. So, just one lesson or activity where AI can build joy where you're incorporating AI generating math games or riddles, belonging, including surveys. Leverage AI for teacher insight, and then it gives me a coaching prompt. So, that's the thing. It's like I know how to generate joy when I was in my own classroom. How do I teach, how do I *coach* my teachers to do the same thing when I don't want them to be like me? I want them to be themselves, authentic to who they are and how they foster that character. This is so cool. |
| 7 | Alexandria | My problem is similar to yours. It said, "I'm trying to hype my students in [inaudible] math. I want to bring joy and fun into the lessons and that's sixth grade students. How do I hype them up and engage them in the math?" And they gave me like, "Build the hype, make the catchy class name, gamify, add drama and storytelling, use student voice and choice, and then build confidence with the growth mindset." And I also prompted after that, I said, "A lot of my students are not confident in doing math and they don't enjoy it. So how do I bring the confidence to math?" And just being the same stuff, like "create the sense of belonging, further confidence gently, praise the process and not the person, make each [inaudible]." |
| 8 | Paule | I'm impressed. |
| 9 | Andres | Yeah. I feel like when you give it a very specific prompt like, "Create a lesson," it was like you said, it focused more on the actual lesson activity part of it and gave me like one or two things. But when you focus on, "This is what I really want to target. How can I do that?" it gave me so many different scenarios that you can go through. Because then I went in after that — I would say, "Give me a specific example of an activity where a student may struggle and prompts on how I can support them." And it was like, "Oh, here's this activity dealing with multi-step word problems with decimals." It's like it literally tells you, "Oh, they might struggle here." Which I think for new teachers especially, it's like you have to identify where the misconceptions are going to be. So, AI for that is like, "Oh, students might do this, this, this." And then gives you, to get them started, like, "Here's the prompts you can use to help organize thinking. What's your first step? Let's just focus on that, to normalize the struggle." It's things that you're not going to think of to say unless you're a veteran in the field. It's like, "That's a tough question." Like if you're telling your kid it's a tough question and they got the first part, that's all the confidence building they might need. So yeah, I think the prompts that you put in definitely get you that output that you bring up. |
| 10 | | [inaudible excerpt] |
| 11 | Nikkie | I just wanted to share one other thing that I really like to do. At the end of my prompt, I say, "I'm a math teacher teaching this grade," similar to how you started. And then I say, "I need you to act like a hype chatbot. Ask me all the questions you need to ask me to generate the best responses for me." I love doing that. I love asking it to take on a role and then making it give me questions. It gave me 14 questions about my students, like "What are they struggling with, what motivates them." So, to my question earlier, do you want to hype up a single student or the whole classroom? I can tell it like, "I have these students, these are their interests" — don't use their real names — but like, these are the interests in my classroom, and then it could give me more specific type scenarios, and then it gave me, like I said, assume things about my classroom, it gave me what all the assumptions are and what would be helpful for those kids. |
| 12 | Andres | That's cool. I love that idea of giving it [a role] because I feel like I do it all the time. I'm like, "Act like a sixth grader. How would you answer this?" And the answers they give are pretty spot on for how kids actually answer something. |
| 13 | Carwai | It's like a [inaudible] thought partner. Like you can make them whoever you need them to be. Like, "Tell me how my 8th grader might react to this particular..." — like giving them a personality. |
| 14 | Nikkie | And so what you started with, you could say, "Act like an administrator. How would you look at this lesson? What would you have questions about?" |
| 15 | Andres | I also find that it gives you ways to word things. "I don't want to sound combative. How can I say this?" — in like, almost like the TikTok memes — but it's like, "Oh, how do you say this professionally?" It's like, but truly, it is really helpful to give you responses and just the tone of how to say it. |

**Appendix Q**
**Post Summit Survey Results**

Post-summit survey responses (n = 54, 89% response rate) suggest that participants had a positive experience at the summit. All respondents rated the summit as productive or very productive (M = 4.57/5), and all reported being likely or very likely to recommend it to a colleague (M = 4.85/5). Among the three summit goals, teachers rated the summit most highly on *Energize* (69% "extremely well"), followed by *Share* (48%), and *Discover* (37%). Facilitators were rated as very or extremely effective by 96% of respondents, and 93% felt the quantity of content was appropriate. All (100%) of respondents expressed interest in attending similar events in the future, 94% were willing to participate in follow-up interviews, and 59% volunteered to serve as community ambassadors, which serve as indicators of sustained engagement. Justice described the summit as "one of the most impactful professional experiences I have engaged in." Teachers' perceptions of AI's impact on math instruction were cautiously optimistic: while 61% rated the expected impact as positive or very positive, 39% remained neutral and none were negative, suggesting that the summit fostered measured hopefulness rather than uncritical enthusiasm. Below we report results for all closed-ended questions.

**Table Q1**
*Overall Ratings*

| Question | Response | n | % |
|---|---|---|---|
| Overall, how productive do you think the summit was for you? | 5- Very Productive | 31 | 57% |
| | 4- Productive | 23 | 43% |
| How likely would you be to recommend this event to a friend or colleague? | 5- Very likely | 46 | 85% |
| | 4- Likely | 8 | 15% |
| What kind of impact do you think AI will have on math instruction? | 5- Very Positive | 6 | 11% |
| | 4- Positive | 27 | 50% |
| | 3- Neutral | 21 | 39% |
| How effective were whole group facilitator(s) in creating a clear, inclusive, and engaging session? | 5- Extremely effective | 31 | 57% |
| | 4- Very effective | 21 | 39% |
| | 3- Moderately effective | 2 | 4% |

Table P2 shows results for how participants rated how well the summit supported three goals. Of 45 respondents, 42 (93%) felt the right amount of content was covered, 2 felt not enough was covered, and 1 felt too much was covered.

**Table Q2**

*Summit Goals*

| Goal | Extremely well | Very well | Moderately well | n |
|---|---|---|---|---|
| Discover | 20 (37%) | 25 (46%) | 9 (17%) | 54 |
| Share | 26 (48%) | 24 (44%) | 4 (7%) | 54 |
| Energize | 37 (69%) | 16 (30%) | 1 (2%) | 54 |

**Table Q3**

*Preparation and financial support.*

| Question | Response | n | % |
|---|---|---|---|
| Enough information to prepare? | 5- Definitely yes | 27 | 50% |
| | 4- Probably yes | 23 | 43% |
| | 3- Might or might not | 4 | 7% |
| Financially supported enough? | 5- Definitely yes | 39 | 72% |
| | 4- Probably yes | 12 | 22% |
| | 3- Might or might not | 2 | 4% |
| | 2- Probably not | 1 | 2% |

**Table Q4**

*Future Engagement*

| Question | Yes | No | n |
|---|---|---|---|
| Interested in attending similar events in the future? | 54 (100%) | 0 (0%) | 54 |
| Willing to participate in a follow-up interview? | 51 (94%) | 3 (6%) | 54 |
| Interested in supporting the research? | 49 (91%) | 5 (9%) | 54 |
| Interested in serving as ambassador/facilitator? | 32 (59%) | 22 (41%) | 54 |